\documentclass[prd,preprint,eqsecnum,nofootinbib,amsmath,amssymb,
               tightenlines,dvips]{revtex4}

\usepackage{graphicx}
\usepackage{bm}

\def\brem{bremsstrahlung}
\def\lra{\leftrightarrow}
\def\x{{\bm x}}
\def\p{{\bm p}}
\def\k{{\bm k}}
\def\q{{\bm q}}
\def\v{{\bm v}}

\def\h{{\bm h}}
\def\u{{\bm u}}
\def\F{{\bm F}}

\def\Q{{\cal Q}}
\def\Re{{\rm Re}}
\def\Im{{\rm Im}}
\def\grad{{\bm\nabla}}
\def\twotwo{{2\lra2}}

\def\half{{\textstyle\frac{1}{2}}}

\def\quarter{{\textstyle\frac{1}{4}}}

\def\PiR{\Pi_{\rm Ret}}
\def\SigmaR{\Sigma_{\rm Ret}}

\def\eff{{\rm eff}}
\def\veps{\varepsilon}

\def\ca{C_{\rm A}}
\def\cf{C_{\rm F}}
\def\da{d_{\rm A}}
\def\df{d_{\rm F}}

\def\Nf{N_{\rm f}}
\def\Nc{N_{\rm c}}
\def\mD{m_{\rm D}}
\def\mg{m_{{\rm eff,g}}}
\def\mf{m_{{\rm eff,f}}}
\def\mq{m_{{\rm eff,q}}}
\def\drangle{\rangle\!\rangle}
\def\dlangle{\langle\!\langle}

\def\Dlangle{\left\langle\!\!\left\langle}
\def\Drangle{\right\rangle\!\!\right\rangle}

\def\n{{\hat{\bm n}}}

\def\LPM{Landau-Pomeranchuk-Migdal}
\def\supLPM{{\scriptscriptstyle \rm LPM}}
\def\Mrow{Mr\'owczy\'nski}
\def\ul{\underline}
\def\dpslash{\frac{d^3\p}{(2\pi)^3} \>}
\def\dprel{\frac{d^3\p}{2|\p| \, (2\pi)^3} \>}
\def\degen{\nu}
\def\meff{m_{\rm eff}}
\def\meffs{m_{{\rm eff},s}}
\def\Tform{t_{\rm form}}
\def\Nform{N_{\rm form}}
\def\Phard{p_{\rm hard}}
\def\Pscreen{p_{\rm screen}}
\def\Pscatter{p_{\rm scatter}}
\def\Pprimary{p_{\rm primary}}
\def\kgamma{k_\gamma}
\def\Kgamma{K_\gamma}
\def\splitsym{\gamma}

\def\gsim{\mbox{~{\protect\raisebox{0.4ex}{$>$}}\hspace{-1.1em}
	{\protect\raisebox{-0.6ex}{$\sim$}}~}}
\def\lsim{\mbox{~{\protect\raisebox{0.4ex}{$<$}}\hspace{-1.1em}
	{\protect\raisebox{-0.6ex}{$\sim$}}~}}

\def\slashchar#1{\setbox0=\hbox{$#1$}           
   \dimen0=\wd0                                 
   \setbox1=\hbox{/} \dimen1=\wd1               
   \ifdim\dimen0>\dimen1                        
      \rlap{\hbox to \dimen0{\hfil/\hfil}}      
      #1                                        
   \else                                        
      \rlap{\hbox to \dimen1{\hfil$#1$\hfil}}   
      /                                         
   \fi}                                         %

\advance\parskip 1.9pt
\advance\voffset -0.2in

\begin{document}

\vspace*{1cm}
\preprint{UW/PT 02-21}

\title{Effective Kinetic Theory for High Temperature Gauge Theories}

\author{Peter Arnold}
\affiliation
    {%
    Department of Physics,
    University of Virginia,
    Charlottesville, Virginia 22901
    }%
\author{Guy D. Moore\footnote{Current address: Department of Physics,
	McGill University, 3600 University St., 
	Montr\'{e}al QC H3A 2T8, Canada}
    and Laurence G. Yaffe}
\affiliation
    {%
    Department of Physics,
    University of Washington,
    Seattle, Washington 98195
    }%

\date{\today}

\begin{abstract}
Quasiparticle dynamics in relativistic plasmas associated with
hot, weakly-coupled gauge theories (such as QCD at asymptotically high
temperature $T$)
can be described by an effective kinetic theory,
valid on sufficiently large time and distance scales.
The appropriate Boltzmann equations
depend on effective scattering rates for various
types of collisions that can occur in the plasma.
The resulting effective kinetic theory may be used to
evaluate observables which are dominantly sensitive to the dynamics
of typical ultrarelativistic excitations.
This includes transport coefficients
(viscosities and diffusion constants) and energy loss rates.
In this paper, we show how to formulate effective Boltzmann
equations which will be adequate to compute such observables
to leading order in the running coupling $g(T)$ of
high-temperature gauge theories [and all orders in $1/\log g(T)^{-1}$].
As previously proposed in the literature,
a leading-order treatment requires including both $2\lra2$
particle scattering processes as well as
effective ``$1\lra2$'' collinear splitting processes
in the Boltzmann equations.
The latter account for
nearly collinear \brem\ and pair production/annihilation processes
which take place in the presence of fluctuations in the background
gauge field.
Our effective kinetic theory is applicable not only to
near-equilibrium systems (relevant
for the calculation of transport coefficients), but also to highly
non-equilibrium situations, provided some
simple conditions on distribution functions are satisfied.
\end{abstract}



\maketitle


\section {Introduction}

In a hot, weakly coupled gauge theory,
such as QCD at asymptotically high temperature where
the running coupling $g(T)$ is small,
one might hope to achieve solid theoretical control
over the dynamics of the theory.
To date, however, very little has been derived about the dynamics of
such theories at even {\it leading order}\/ in the coupling ---
that is, neglecting all relative corrections to the leading
weak-coupling behavior which are suppressed by powers of $g$.
For example, hydrodynamic transport properties such as shear
viscosity, electrical conductivity, and flavor diffusion are not
known at leading order; they have only
been calculated in a ``leading-log'' approximation,
which has relative errors of order
$1/\log(g^{-1})$.%
\footnote{
  Ref.\ \cite{heiselberg} attempted to calculate the shear viscosity
  to next-to-leading logarithmic order
  [including relative corrections of order $1/\log(g^{-1})$
  but neglecting $1/\log^2(g^{-1})$ effects]
  but, among other things, missed the ``$1\lra2$'' collinear
  processes described in this paper, as well as pair annihilation/creation
  processes which contribute even at leading-log order.
  The latter have been previously discussed in Ref.\ \cite{leadinglog}.
}%

To study transport or equilibration processes
in a hot plasma quantitatively,
the most efficient approach is first to construct an
effective kinetic theory which reproduces,
to the required level of precision,
the relevant dynamics of the underlying quantum field theory,
and then apply this kinetic theory to the processes of interest.
Specifically, we would like to formulate an appropriate set of
Boltzmann equations which will, on sufficiently long time and distance
scales, correctly describe the dynamics of typical ultrarelativistic
excitations ({\em i.e.}, quarks and gluons) with sufficient accuracy
to permit a correct leading-order evaluation of observables such
as transport coefficients.
Schematically, these Boltzmann equations will have the
usual form,
\begin {equation}
   ( \partial_t + \v\cdot\grad_\x ) \, f = - C[f] ,
\label {eq:boltz}
\end {equation}
where $f = f(\x,\p,t)$ represents the phase space density of (quasi-)particles
at time $t$,
$\v$ is the velocity associated with an excitation of momentum $\p$,
and $C[f]$ is a spatially-local collision term that represents the rate
at which particles get scattered out of the momentum state $\p$ minus
the rate at which they get scattered into this state.
The challenge is to understand exactly what processes need to be
included in the collision operator $C[f]$, and how to package them,
so that the Boltzmann equation correctly reproduces the desired physics
at the required level of accuracy.

To compute transport coefficients or asymptotic equilibration rates,
one does not actually need a general non-equilibrium (and non-linear)
kinetic theory;
it is sufficient merely to have Boltzmann equations linearized in small
deviations away from an equilibrium state of given temperature $T$.%
\footnote 
    {%
    See, for example, the discussion in Ref.~\cite {leadinglog}.
    }
But more generally one would like to formulate a fully
non-equilibrium kinetic theory which would also be applicable to systems
(such as intermediate stages of a heavy ion collision \cite {BMSS})
in which deviations from equilibrium are substantial and quantities such
as temperature are not unambiguously defined.
This will be our goal.
It should be emphasized that all of our analysis assumes
the theory is weakly coupled on the scales of interest.
Hence, the domain of validity includes QCD
at asymptotically high temperatures, or intermediate stages of
collisions between arbitrarily large nuclei at asymptotically
high energies \cite {BMSS}.
It is an open question to what extent
this weak coupling analysis is applicable to real heavy ion collisions
at accessible energies.
However, we believe that understanding the dynamics in
weakly coupled asymptotic regimes is a necessary and useful
prerequisite to understanding dynamics in more strongly coupled regimes.

The domain of applicability of any kinetic theory depends on the
time scales of the underlying scattering processes which are
approximated as instantaneous transitions in the collision term
of the Boltzmann equation.
In the remainder of this introduction,
we review the relevant scattering processes and associated time scales,
describe the assumptions underlying
our effective kinetic theory for near-equilibrium systems
in which the temperature is at least locally well-defined,
and then discuss how the required conditions can be generalized
to a much wider class of non-equilibrium systems.


\subsection {Relevant scattering processes}\label{sec:scales}

Consider a QCD plasma at sufficiently high temperature $T$ so that
the effective coupling $g(T)$ is small.%
\footnote
    {%
    For convenience of presentation,
    we will use the language of QCD throughout this paper.
    Despite this, all our discussion and results apply equally well to
    high temperature electroweak theory and,
    except for a few comments about non-perturbative
    non-Abelian $g^2 T$ scale physics, also to high temperature QED.
    }
Quarks and gluons are well-defined quasiparticles of this system.
The typical momentum (or energy) of a quark or gluon is of order $T$;
this will be referred to as ``hard.''
The number density of either type of excitation is $O(T^3)$,
so that the total energy density is $O(T^4)$.
Hard quarks and gluons propagate as nearly free, nearly massless excitations.
Their dispersion relations receive thermal
corrections which look just like an effective mass,%
\footnote
    {%
    This simple form holds, up to yet higher-order corrections,
    provided the momenta of the excitation is large compared to the
    thermal mass, $|\p| \gg O(gT)$.
    Our $m_{\rm th}$ is often called the asymptotic thermal mass.
    }
\begin {equation}
    \epsilon(\p) = \sqrt {\p^2 + m_{\rm th}^2} \,,
\end {equation}
with the thermal masses for both quarks and gluons being $O(gT)$ in size.
Hence, all but a parametrically small $O(g^2)$ fraction of excitations
[those with $O(gT)$ momenta] are ultrarelativistic
and travel at essentially the speed of light.
If $\hat\p$ denotes the unit vector in the direction of $\p$ then,
for hard excitations,
\begin {equation}
    \v(\p) \equiv \frac {\partial \epsilon(\p)}{\partial\p}
    = \hat\p \, [1 - O(g^2)] \,.
\end {equation}

\begin{figure}
  \includegraphics[scale=0.30]{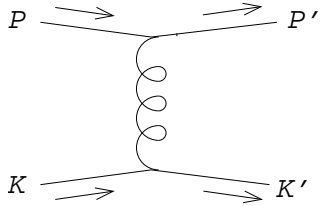}
  \caption{
    Scattering of hard particles by $t$-channel gauge boson exchange.
    The straight lines could represent any type of hard charged particle,
    including fermions, gauge bosons, or scalars.
    \label{fig:tchannel}
  }
\end{figure}

Scattering processes in the plasma will cause any excitation to have
a finite lifetime.
Imagine focusing attention on some particular excitation with
a hard $O(T)$ momentum.
What determines the fate of this quasiparticle?
One relevant process is ordinary Coulomb scattering,
depicted in Fig.~\ref{fig:tchannel}.
If the particle of interest scatters off some other excitation in the plasma
with a momentum transfer $\q$,
then the direction of the particle can change by
an angle $\theta$ which is $O(|\q|/T)$.
The differential scattering rate is
\begin {equation}
    d\Gamma
    \sim g^4 \, T^3 \, \frac {dq}{q^3}
    \sim g^4 \, T \, \frac {d\theta}{\theta^3} \,.
\end {equation}
This form holds provided $q \gsim O(gT)$.
Below this scale, Debye screening (and Landau damping)
in the plasma soften the small angle divergence of the
bare Coulomb interaction.

Consequently, the rate for a single large angle scattering
with $O(T)$ momentum transfer is $O(g^4 T)$,
while the rate for small angle scattering with $O(gT)$
momentum transfer is $O(g^2 T)$.
For later use, let $\tau_g = 1/(g^2 T)$ denote the characteristic
small angle mean free time,
and $\tau_* = 1/(g^4 T)$ the characteristic large angle mean free time,%
\footnote
    {%
    \label {fn:logs}%
    Actually, because small angle scatterings are individually more
    probable than large angle scatterings, a particle can undergo
    $N_g \sim 1/g^2$ small angle scatterings,
    each with $\theta = O(g)$, during a time of order
    $N_g \, \tau_g \sim 1/(g^4 T)$ --- the same as the
    time for a single $q = O(T)$ scattering.
    A succession of this many uncorrelated small angle scatterings
    will result in a net deflection of order $N_g^{1/2} \, \theta = O(1)$.
    Hence, a large deflection in the direction of a particle
    is equally likely to be the result of many small angle scatterings
    or a single large angle scattering.
    In fact, the multiplicity of possible combinations of scatterings
    with angles between $\theta \sim g$ and $\theta \sim 1$ leads to
    a logarithmic enhancement in the large angle scattering rate,
    or a logarithmic decrease of the large angle mean free path,
    so that $\tau_* \sim [g^4 T \log(1/g)]^{-1}$.
    In this paper, we will ignore such logarithmic factors when
    making parametric estimates and simply write $\tau_* \sim 1/g^4 T$.
    Nevertheless, it is important to keep
    in mind that contributions to $\tau_*$ come from the
    entire range of scatterings from $\theta \sim g$ to $\theta \sim 1$,
    which corresponds to momentum transfers from $q \sim gT$ to $q \sim T$.
    \\[4pt]
    Yet softer momentum transfers are screened in the plasma
    and do not affect $\tau_*$ at the order of interest.
    For instance, the mean free path for very-small angle scattering
    with $\theta\sim g^2$ (corresponding to momentum transfers of order
    $g^2 T$) is only $\tau_{g^2} \sim 1/g^2 T$ and is not enhanced over
    $\tau_g$.
    Over the time $\tau_* \sim 1/g^4 T$, there can thus be only
    $N_{g^2} \sim 1/g^2$ independent very-soft scatterings,
    which will only contribute a net deflection of order
    $\Delta\theta \sim (N_{g^2})^{1/2} g^2 \sim g$
    to the $O(1)$ deflection caused by other processes.
    Consequently, the large angle scattering time
    $\tau_*$ is insensitive to
    non-perturbative magnetic physics in the plasma
    associated with very soft momentum transfers of order $g^2T$.
    }
also known as the transport mean free time.
Neither time has a precise, quantitative definition;
these quantities will only be used in parametric estimates.
The small angle mean free time $\tau_g$ is
[to within a factor of $O(\log g^{-1})$]
the same as the color coherence time of an excitation ---
this is the longest time scale over which it makes sense to think
of an excitation has having a definite (non-Abelian) color
\cite {Blog1,Selikhov&Gyulassy,Bodeker98,Bodeker99}.
These scales are summarized in Table~\ref{tab:scales}.

\begin {table}
    \begin {center}
    \tabcolsep 10pt
    \begin {tabular}{lll}
    \hline 
       typical particle wavelength               & $T^{-1\strut}$   \\
       typical inter-particle separation         & $T^{-1}$         \\[3 pt]
    \hline
       Debye screening length                    & $(gT)^{-1\strut}$      \\
       inverse thermal mass                      & $(gT)^{-1}$      \\[3 pt]
    \hline
       mean free path: small-angle scattering ($\theta\sim g$, $q\sim gT$)
                                                 & $(g^2 T)^{-1\strut}$   \\
       mean free path: very-small-angle scattering
             ($\theta\sim g^2$, $q\sim g^2T$)
                                                 & $(g^2 T)^{-1}$   \\
       duration (formation time) of ``$1\lra2$'' collinear processes
                                                 & $(g^2 T)^{-1}$   \\
       non-perturbative magnetic length	scale for colored fluctuations
                                                 & $(g^2 T)^{-1}$   \\[3pt]
    \hline
       mean free path: large-angle scattering ($\theta\sim1$, $q\sim T$)
                                                 & $(g^4 T)^{-1\strut}$   \\
       mean free path: hard ``$1\lra2$'' collinear processes
                                                 & $(g^4 T)^{-1}$   \\[3 pt]
    \hline
    \end {tabular}
    \vspace*{-5pt}
    \end {center}
    \caption
        {%
        Parametric dependence of various length scales for a weakly-coupled
        ultrarelativistic equilibrium plasma.
        Estimates for mean free paths apply to typical
	(hard) particles,
        with $\theta$ and $q$ denoting the deflection angle
        and momentum transfer, respectively.
        The non-perturbative magnetic physics scale of $(g^2 T)^{-1}$ for
        colored fluctuations applies only to non-Abelian gauge theories.
        }
    \label {tab:scales}
    \medskip
\end {table}

\begin{figure}
  \includegraphics[scale=0.70]{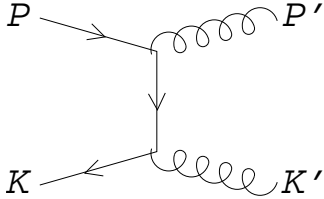}
  \vspace*{-5pt}
  \caption{
    $t$-channel diagram for $q\bar q\to gg$.
    \label{fig:qqgg}
  }
\end{figure}

It is also important to consider processes which change
the {\it type}\ of an excitation.
Consider, for example, the conversion of a quark of momentum $\p$
into a gluon of nearly the same momentum by the
soft $q \bar q \to g g$ process,
depicted in Fig.~\ref{fig:qqgg},
with momentum transfer $\q \sim g T$.
The mean free path for this process (or its time-reverse) is $O[1/(g^4 T)]$,
just like the large angle scattering time $\tau_*$.
As a result,
such $t$-channel quark exchange processes
are equally important as gluon exchange for quasiparticle dynamics,
and must be correctly included even in leading-log evaluations of
transport coefficients~\cite{leadinglog}.

Crossed $s$-channel versions of Figs.~\ref{fig:tchannel} and \ref{fig:qqgg},
namely quark-antiquark annihilation and creation via a single virtual gluon,
and gluo-Compton scattering,
also proceed at $O(g^4 T)$ rates.
Consequently, these processes must also be included in a leading-order
treatment of quasiparticle dynamics.%
\footnote
    {%
    These $s$-channel processes
    do not have the logarithmic enhancement mentioned in
    footnote \ref{fn:logs},
    and so
    do not contribute to transport coefficients at leading-log order,
    but do contribute at next-to-leading log order.
    }
Henceforth, whenever we refer to $2\lra2$ particle processes,
we will mean all possible crossings of Figs.~\ref{fig:tchannel}
and \ref{fig:qqgg} in which two excitations turn into two excitations.

\begin{figure}
  \includegraphics[scale=0.60]{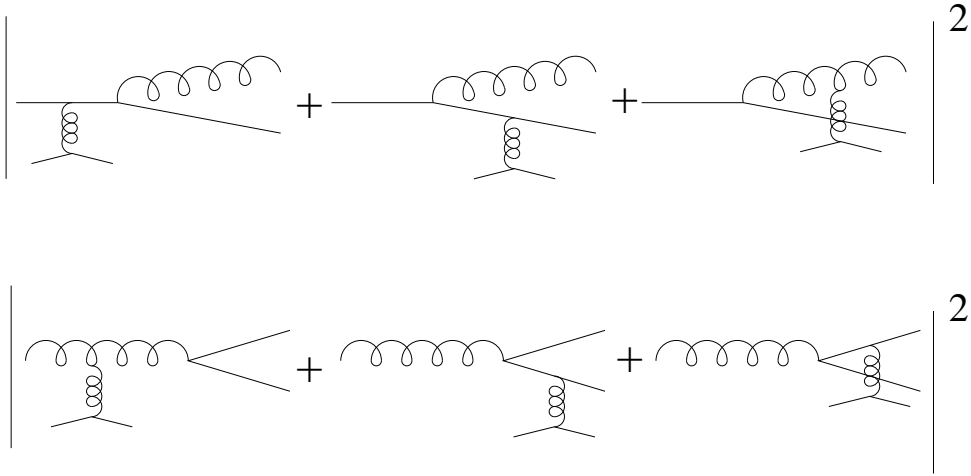}
  \caption{
    Simplest examples of near-collinear processes which contribute
    at leading order.
    The upper diagrams depict hard, nearly collinear gluon \brem\
    accompanying a soft gluon exchange between hard particles.
    The lower diagrams show the conversion of a hard gluon into a 
    nearly collinear quark-antiquark pair when accompanied
    by a soft exchange with another excitation in the plasma.
    \label{fig:brem}
  }
\end{figure}

In addition to the $2\lra2$ particle processes of Figs.~\ref{fig:tchannel}
and \ref{fig:qqgg},
hard quasiparticles in the plasma can also undergo processes in which they
effectively split into two different, nearly collinear, hard particles.
Such processes cannot occur (due to energy-momentum conservation)
in vacuum, but they become kinematically allowed when combined with
a soft exchange involving some other excitation in the plasma.
A specific example is the \brem\ process depicted in the upper
part of Fig.\ \ref{fig:brem}.
Here, one hard quark undergoes
a soft ($\q \sim gT$) collision with another and splits into a hard quark
plus a hard gluon,
each of which carry an $O(1)$ fraction of the hard momentum of the
original quark and both of which travel in
the same direction as the original quark to within an angle of $O(g)$.
The mean free path for this process, as well as the
near-collinear pair production process also shown in Fig.~\ref{fig:brem},
turns out to be $O[1/(g^4 T)]$, which is once again the same order
as the large angle scattering time $\tau_*$.%
\footnote
    {
    See, for example, Ref.~\cite {sansra} as well as the
    discussion of the closely related case of photon emission in
    Refs.~\cite {powercount,Gelis1}.
    More careful analysis shows that the rates of these
    near-collinear processes do not have the logarithmic 
    enhancement of the large angle scattering rate
    discussed in footnote \ref {fn:logs}.
    Therefore these near-collinear processes do not
    contribute to transport coefficients at leading-log order
    but must be included at next-to-leading log order.
    }

\begin{figure}
  \includegraphics[scale=0.40]{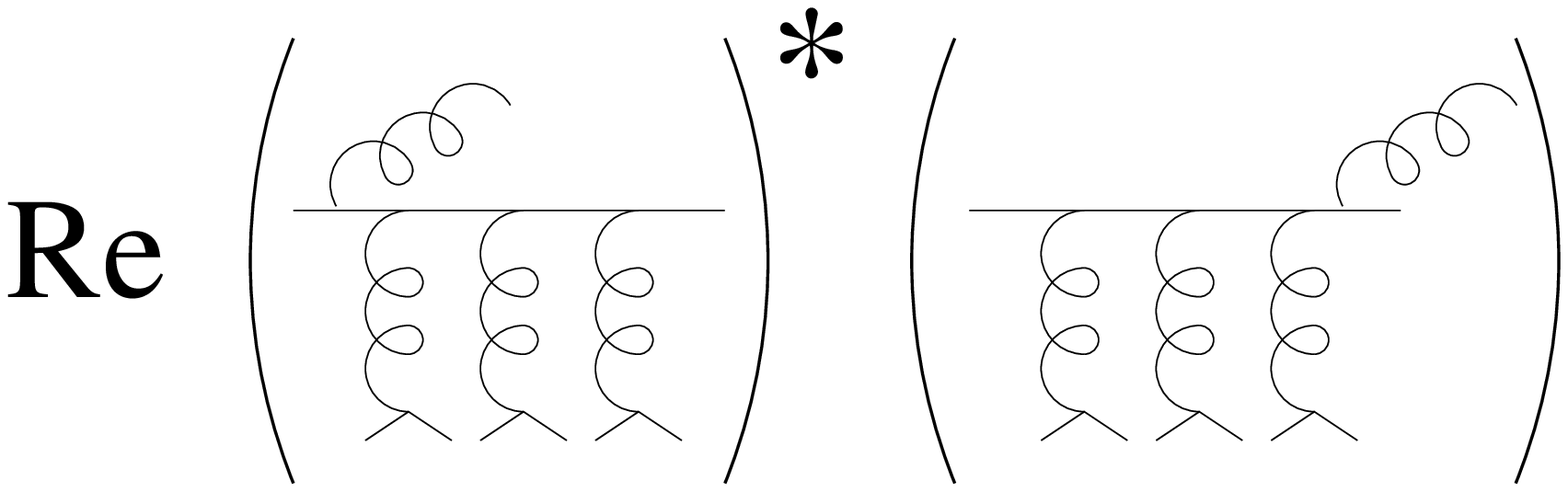}
  \caption{
    Examples of interference terms involving multiple scatterings that
    also contribute to hard-gluon \brem\ at leading order.
    \label{fig:lpm}
  }
\end{figure}

There is an important difference between these near-collinear processes and
the $2\lra2$ particle processes of Figs.~\ref{fig:tchannel} and \ref{fig:qqgg}.
The intermediate propagator in the diagrams of Fig.~\ref {fig:brem} which
connects the ``splitting'' vertex with
the soft exchange has a small $O(g^2T^2)$ virtuality.
Physically, this means that the time duration of the near-collinear processes
(also known as the scattering time or formation time)
is of order $1/g^2 T$, which is the same as the
mean free time $\tau_g$ for small-angle elastic collisions in the plasma.
As a result, additional soft collisions are likely to occur during the
splitting process and can disrupt the coherence between the nearly
collinear excitations.
This is known as the \LPM\ (LPM) effect.
Since the time between soft scatterings is comparable to the
time duration of the emission process,
multiple soft scatterings cannot be treated as independent classical events,
but must be evaluated fully quantum mechanically.
In other words, including the interference between different
$N+1 \to N+2$ amplitudes,
such as depicted in Fig.\ \ref{fig:lpm}, is required to
evaluate correctly the rate for near-collinear splitting at leading order.
We will refer to these processes collectively as ``$1\to2$'' processes
where the $1 \to 2$ refers to the nearly collinear splitting
particles and the quotes are a reminder that there are other
hard particles participating in the process via multiple soft gluon exchanges.
Of course, the inverse ``$2 \to 1$'' processes,
in which two nearly collinear excitations fuse into one,
are also required for detailed balance.
The evaluation of the rate of these ``$1\lra2$'' near-collinear processes
in an equilibrium ultrarelativistic plasma, complete to leading order,
is discussed in Ref.\ \cite{sansra},%
\footnote{
  See also Refs.~\cite {powercount,photo_emit} and references therein.
}
which derives a two-dimensional linear integral equation whose
solution determines the leading order rate.

Having realized that near-collinear ``$1\lra2$'' processes
are just as important to the fate of a quasiparticle as $2\lra2$
particle processes,
one might wonder if there are any further relevant processes which occur
at an $O(g^4 T)$ rate and hence compete with the processes just discussed.
We will argue in section \ref{sec:addl-processes} that this is not the case.

In summary, a typical hard excitation travels a distance of order
$\tau_* \sim 1/(g^4T)$ before it either experiences a large angle scattering,
converts into another type of excitation, splits into two
nearly collinear hard excitations, or merges with another nearly
collinear excitation.
These different ``fates'' are all comparably likely
[up to factors of $O(\log g^{-1})$].
During the transport mean free time $\tau_*$,
the excitation will experience many
soft scatterings each of which can completely reorient the
color of the quasiparticle,
but only change its momentum by a small $O(gT)$ amount.

\subsection {Kinetic theory domain of validity --- near-equilibrium systems}

A near-equilibrium system is one in which the phase space
distribution can be written as an equilibrium
distribution $n(\p)$ plus a small perturbation $\delta f(\x,\p)$.
However, we wish to include situations where the system is close to
local rather than global equilibrium, so that equilibrium parameters
like temperature may vary slowly with $\x$.
We will therefore define
a near-equilibrium system to be one where
distribution functions can be written in the form
\begin {equation}
   f(\x,\p) = n\bigl(\p; T(\x), \u(\x), \mu(\x)\bigr) + \delta f(\x,\p) ,
\label {eq:near_equilibrium}
\end {equation}
where $n(\p; T, \u, \mu)$ denotes an equilibrium distribution
(Bose or Fermi, as appropriate) with temperature $T$, flow velocity $\u$, and
optionally one or more chemical potentials $\mu$,
provided:
(i)
the parameters $T(\x)$, $\u(\x)$, and $\mu(\x)$
do not vary significantly over distances (or times) of order $1/g^4 T$,
the large angle mean free path,%
\footnote{
   This condition should be understood as applying in a frame
   which is slowly moving relative to the local rest frame of the
   fluid at point $\x$.
}
and (ii) throughout the system, $\delta f \ll f$.%
\footnote
    {%
    More precisely, $\delta f$ must make a negligible contribution
    to the integrals (\ref {eq:J}) and (\ref {eq:I}) below.
    This ensures that the local-equilibrium part of the distribution
    dominates screening and soft scattering processes.
    }

In the case of near-equilibrium systems,
we can now summarize the conditions upon which
our effective kinetic theory is based.

\begin {enumerate}
\item
    As already indicated, we assume that the theory is weakly coupled
    on the scale of the temperature, $g(T) \ll 1$.
    Consequently, there is a parametrically large separation
    between the different scales shown in Table~\ref{tab:scales}.

\item
    We assume that all zero-temperature mass scales
    ({\em i.e.}, $\Lambda_{\rm QCD}$ and current quark masses)
    are negligible compared to the $O(gT)$ scale of thermal masses.%
    \footnote
	{
	For hot electroweak theory, the required condition is that
	the Higgs condensate be small compared to the temperature,
	$v(T) \ll T$,
	so that the condensate-induced mass of the
	$W$-boson is negligible compared to its thermal mass.
	}

\item
    Any kinetic theory can only be valid on time scales
    which are large compared to the duration of the scattering processes
    which are approximated as instantaneous inside the collision term
    of the Boltzmann equation.
    Since the formation time of near-collinear splitting processes
    is order $1/(g^2 T)$, this means we must assume that
    the space-time variation of the deviation from local equilibrium
    $\delta f(\x,\p)$
    is small on the scale of $1/(g^2 T)$.
    [Note that we've already assumed a stronger condition on the
    spacetime variation of the local equilibrium part of distributions,
    $n(\p;T(\x),\u(\x),\mu(\x))$.]

\item
    We will assume that hard particle distribution functions $\delta f(\x,\p)$
    have smooth dependence on momentum $\p$ and do not vary significantly with
    changes in momentum comparable to the $O(gT)$ size of thermal masses.
    [Again, this is automatic for the local equilibrium part of
    distributions.]
    This condition will allow us to simplify our treatment of
    distribution functions for near-collinear
    ``$1\lra2$'' processes.

\item
    Finally, we assume that the observables of ultimate interest
    are dominantly sensitive to the dynamics of hard excitations,
    are smooth functions of the momenta of these excitations,
    do not depend on the spin of excitations,
    and are gauge invariant.
    Excluding spin-dependent observables is primarily a matter of convenience,
    and will allow us to use spin-averaged effective scattering rates.%
    \footnote
	{%
	If this condition is not met, distribution functions become
	density matrices in spin space, and the Boltzmann equation
	must be replaced by a generalization known as a
	Waldmann-Snider equation \cite {Snider}.
	(See also the discussion in Ref.~\cite {Blog1} for the analogous 
	case of color-dependent density matrices.)
	We assume spin-independence to avoid needlessly complicating
	this paper,
	and because we are not aware of physically interesting problems
	in hot gauge theories where strong spin polarization is relevant.
	}
    Considering color-dependent observables would be senseless
    in an effective theory which is only applicable on spatial scales
    large compared to the $1/(g^2 T)$ color coherence length.

\end {enumerate}


\subsection {Kinetic theory domain of validity --- non-equilibrium case}
\label {sec:non-eq}

It is impossible for any effective kinetic theory
to be valid for all possible choices of phase space distribution functions.
There will always exist sufficiently pathological initial states
which are outside the domain of applicability of any kinetic theory.
Describing the domain of applicability therefore requires
a suitable characterization of acceptable distribution functions.
In a general non-equilibrium setting, we will define
``reasonable'' distribution functions to be those which
support a separation of scales similar to
the weakly-coupled equilibrium case.
That is, the momenta of relevant excitations must be
parametrically large compared to medium-dependent
corrections to dispersion relations or to the
inverse Debye screening length.
Furthermore, the phase space density of excitations
must not be so large as to drive the dynamics,
on these scales, into the non-perturbative regime.

There are potentially many more relevant scales one may need
to distinguish in a non-equilibrium setting.
In particular, in place of the single hard scale $T$
that characterizes the momenta of typical excitations in equilibrium,
one may need to consider at least three relevant scales,
which we shall refer to
as the typical momenta of (i) primaries, (ii) screeners, and (iii) scatterers.
By ``primaries,'' we mean the particles whose evolution we are explicitly
interested in following with the Boltzmann equation.  These might be
the particles which dominate the energy of the system (as in the intermediate
stages of bottom-up thermalization \cite{BMSS}), they might be particles which
dominate transport of some charge of the system we are following, or whatever.
These excitations may or may not overlap with
the next two categories.  The ``screeners'' are those
particles whose response to electric and magnetic fields dominates the
screening effects and hard thermal masses in the medium.  That is, they
have momenta comparable to the momentum scale at which
hard thermal-loop (HTL) self-energies
receive their dominant contribution.
(Explicit formulas for HTL self-energies will be reviewed in
Section \ref{sec:self-energies}.)
Finally, the
``scatterers'' are those particles which generate the soft background
of gauge fields off of which the primaries scatter in soft collisions.
As a pictorial example, consider
Figs.~\ref{fig:tchannel}, \ref{fig:brem}, and \ref{fig:lpm}.
The lines entering on the top are primaries,
the lines entering on the bottom are scatterers, and the screeners
are not shown explicitly but are the particles in the hard thermal
loops that are implicitly summed in the soft, exchanged gluon lines.

To formulate the conditions under which our Boltzmann equations
will be valid, 
it will be helpful
to define the following two integrals involving the distribution function
of a particular species (gluon, quark, or antiquark) of quasiparticles,
\begin {eqnarray}
    {\cal J}(\x) &\equiv& \int \dpslash \, \frac{f(\p,\x)}{|\p|} \,.
\label {eq:J}
\\[5pt]
    {\cal I}(\x) &\equiv&
    \half \int \dpslash \, f(\p,\x) \, [1 \pm f(\p,\x)] \,,
\label {eq:I}
\end {eqnarray}
As usual upper signs refer to bosons and lower signs to fermions.
In equilibrium, the ratio ${\cal I} / {\cal J}$ is precisely the temperature
$T$.

Corrections to the ultrarelativistic dispersion relation
for quasiparticles will be seen to involve the quantity ${\cal J}(\x)$;
the square of medium-dependent effective masses will be of order
$g^2 \, {\cal J}$.
So the momentum dominating this integral defines the typical momentum
of screeners which we will denote by $\Pscreen$,
and
\begin {equation}
    \meff \sim g \sqrt {\cal J}
\label {eq:meff}
\end {equation}
is the characteristic size of effective masses.
This reduces to the $gT$ scale in the near-equilibrium case.
Momenta large compared to $\meff$ will be referred to as ``hard.''

The integral ${\cal I}(\x)$ will appear as an effective density
of scatterers which generate soft background gauge field fluctuations.
The momentum scale which dominates this integral is,
by definition, the characteristic momentum of scatterers,
$\Pscatter$.
The mean free time between
small-angle scatterings (with momentum transfer of order $\meff$)
scales as%
\footnote
    {
    Initial and final state factors for the scatterer appear in this
    expression inside the integral ${\cal I}$, given by Eq.~(\ref{eq:I}).
    Some readers may wonder whether there should be analogous initial and
    final state factors for the primary excitation.
    To be more precise, $\tau_{\rm soft}$ represents the inverse of the
    contribution to the thermal width of quasiparticles due to soft scattering.
    This part of the width does not contain
    statistical factors for the primaries.
    The scale $\tau_{\rm soft}$ characterizes the time scale for
    the decay of quasiparticle excitations.
    Equivalently, it is the relaxation time of fluctuations
    in the occupancy of a hard mode due to soft scattering.
    }
\begin {equation}
    \tau_{\rm soft} \sim \frac {\meff^2}{g^4 \, \cal I} \,.
\label {eq:tau_soft}
\end {equation}
This generalizes the $1/(g^2 T)$ small angle mean free time in the
near-equilibrium case.

Let $\Pprimary$ denote the characteristic primary momenta of interest,
and let
$\Phard$ denote the minimum of $\Pscatter$, $\Pscreen$, and $\Pprimary$.
As we shall review in section \ref{sec:tform}, the formation time of
near-collinear bremsstrahlung or annihilation
processes involving a primary excitation behaves
(up to logarithms) as
\begin {equation}
   \Tform \sim \frac{\Pprimary}{\Nform \, \meff^2} \,,
\label {eq:Tform}
\end {equation}
where
\begin {equation}
   \Nform \sim
   \left[1 + \frac{\Pprimary}{\meff^2 \,\tau_{\rm soft}} \right]^{1/2}
\label {eq:Nform}
\end {equation}
represents the typical number of soft collisions occurring during a single
near-collinear ``$1\lra2$'' process.
(In other words, $\Nform$ is parametrically either 1 or
$[\Pprimary/(\meff^2 \,\tau_{\rm soft})]^{1/2}$, depending on whether
$\Pprimary$ is small or large compared to
$\meff^2 \, \tau_{\rm soft}$.)

Since distribution functions may vary in space or time,
all these scales are really local and may vary from point
to point in spacetime.
However, we will assume that distribution functions are
suitably slowly varying in space and time.
The following discussion should be understood as applying
in the vicinity of any particular point $\x$ in the system.

\smallskip

The key assumptions we will make are:
\begin {enumerate}\advance\itemsep 2pt
\item
    The species dependence of the above scales
    is not parametrically large.
    In particular, ratios of the effective masses of different species
    are $O(1)$.
    This assumption is a matter of convenience,
    but relaxing it would significantly complicate the discussion.

\item \label {item:scale-sep}
    The momenta of primaries, scatterers, and screeners are all large
    compared to medium-dependent effective masses.
    A large separation of scales is essential to our analysis,
    and implies that all relevant excitations are highly relativistic.
    To make parametric estimates, we will assume that
\begin {equation}
           \meff / \Phard \lesssim O(g^\alpha)
\end {equation}
    for some positive exponent~$\alpha$.  [For instance, $\meff/ \Phard$
    was $O(g)$ in our previous discussion of near-equilibrium physics.]

\item \label {item:meff}
    The effective mass $\meff$ must be large compared to the
    small-angle scattering rate $\tau_{\rm soft}^{-1}$,
    as well as to zero-temperature mass scales
    ($\Lambda_{\rm QCD}$ and quark masses).

\item
    All distribution functions have negligible variation
    over spacetime regions whose size equals the formation time 
    $\Tform$ of near-collinear processes involving primary excitations.
    This assumption is essential for our treatment of near-collinear
    processes.

\item
    Distribution functions of scatterers and screeners
    do not vary significantly with parametrically small changes
    in the direction of propagation of excitations.
    For example, distributions cannot be so highly anisotropic that
    the directions of screeners or scatterers all lie within an $O(g)$
    angle of each other.
    This simplifies the analysis by allowing
    us to ignore angular dependence when making parametric estimates.
    We will actually find that a stronger limit on the
    anisotropy of screeners is needed in order to prevent the appearance
    of soft gauge field instabilities \cite {Mrow_instability}
    whose growth can lead to violations of the preceding spatial
    smoothness condition.
    Discussion of such instabilities will be postponed to
    section \ref {sec:instability}.

\item
    Distribution functions, for hard momenta,
    have smooth dependence on momentum
    and do not vary significantly with $O(\meff)$
    changes in momentum.
    [This assumption is implicit in various non-equilibrium HTL
    results which we will take from the literature.]
    For momenta of order $\Pprimary$, we further assume
    that distributions do not vary significantly with an
    $O(\Nform^{1/2} \, \meff)$ change in momentum, which represents
    the typical total momentum transfer due to the $\Nform$ soft collisions
    occurring during a near-collinear ``$1\lra2$'' process.

\item\label {item:non-pert}
    Distribution functions are not non-perturbatively large
    for momenta $p \gtrsim O(\meff)$.
    Specifically, for bosonic species we will assume that
    $f(\p,\x)$ is parametrically small compared to $1/g^{2}$.
    (Fermionic distributions can never be parametrically large.)
    As discussed below, this inequality is actually a consequence
    of condition \ref {item:scale-sep}.

\item
    Distribution functions are not spin-polarized.
    Once again, this condition is a matter of convenience.
    But the consistency of this assumption is now a non-trivial issue
    since we are not requiring distribution functions to be isotropic
    in momentum space.
    In a general (rotationally invariant) theory, it is quite possible
    for a medium with anisotropic distribution functions to generate
    medium-dependent self-energy corrections which lead to spin
    or polarization dependent dispersion relations,
    implying spin-dependent propagation of quasiparticles.
    In such a theory, initially unpolarized distribution functions
    would not remain unpolarized.
    This point will be discussed further in section \ref {sec:self-energies},
    where we will see that in hot gauge theories
    (at the level of precision relevant for our leading-order kinetic theory)
    unpolarized but anisotropic distribution functions do not generate
    birefringent quasiparticle dispersion relations.

\item
    Distribution functions are color singlets.
    Attempting to incorporate colored distribution functions
    would be inconsistent, since the color coherence time is comparable
    or shorter than the formation time of the ``$1\lra2$''
    processes which will be treated as instantaneous in this effective
    kinetic theory.

\item
    Observables of interest are dominantly sensitive to the dynamics
    of excitations with momenta of order $\Pprimary$,
    are smooth functions on phase space, are gauge invariant,
    and are spin independent.

\end {enumerate}
The most important restrictions are the scale-separation condition
\ref {item:scale-sep} and the effective mass condition \ref {item:meff}.
They have numerous consequences, including condition~\ref{item:non-pert},
as explained below.

Let $\bar f_p$ denote the average phase space density for
excitations with momenta of order $p$.
First note that the definition of
$\Pscreen$ as the momentum which dominates the integral $\cal J$
(\ref{eq:J}) implies that
\begin {equation}
    p^2 \, \bar f_p \lsim \Pscreen^2 \, \bar f_{\Pscreen} \sim {\cal J},
\label{eq:screen_inequality}
\end {equation}
for any momentum $p$.
For momenta $p$ which are parametrically different from $\Pscreen$,
the above inequality ($\lsim$) is actually a strong inequality $(\ll)$
(or else $\Pscreen$ will not be the scale which dominates $\cal J$).
{}From (\ref {eq:screen_inequality}) and definition (\ref {eq:meff}),
one has
\begin {equation}
    p^2 \, \bar f_p \lsim \meff^2/g^2 .
\label {eq:plop}
\end {equation}
For hard momenta $p \gtrsim \Phard$,
condition~\ref {item:scale-sep}
[$\meff^2/p^2 \lsim O(g^{2\alpha})$]
thus implies that
\begin {equation}
    \bar f_p \lsim O(g^{2\alpha -2}) ,
\label {eq:strongf}
\end {equation}
which is a strengthened version of condition~\ref {item:non-pert}.
For soft momenta, such as $p \sim \meff$, which are parametrically smaller
than $\Pscatter$ by condition~\ref{item:scale-sep},
the inequality (\ref{eq:plop}) becomes strong,
$p^2 \, \bar f_p \ll \meff^2/g^2$.  We can then generally conclude
that for momenta $p \gtrsim \meff$, the phase space density is perturbative,
\begin {equation}
   f_p \ll O(g^{-2}) \,.
\end {equation}

In order for any Boltzmann equation to be valid, the duration of
scattering events (which are treated as instantaneous in the Boltzmann
equation) must be small compared to the typical time between scatterings.
For $2\lra2$ collisions, the largest relevant scattering duration
is $1/\meff$ [for soft scatterings with $O(\meff)$ momentum transfer]
and the smallest relevant mean free time is
the small-angle scattering time (\ref {eq:tau_soft}).
Hence, the condition \ref {item:meff} requirement that
$ \meff \, \tau_{\rm soft} \gg 1 $ is needed for the
validity of kinetic theory.
If the density of scatterers is not parametrically small,
so that $\bar f_{\Pscatter} [1 \pm \bar f_{\Pscatter}]$ is $O(1)$ or larger,
then this inequality automatically holds since
Eqs.~(\ref{eq:screen_inequality}) and (\ref{eq:strongf}) imply%
\footnote
    {%
    This estimate assumes that some of the relevant scatterers
    are bosons, which might have parametrically large distributions.
    If the only relevant scatterers are fermions, then one finds
    $\meff \, \tau_{\rm soft} \gsim O(1/g)$.
    }

\begin {equation}
	\meff \, \tau_{\rm soft}
    \sim
	\frac{{\cal J}^{3/2}}{g \, {\cal I}}
    \sim
	\frac{\bar f_{\Pscreen}^{3/2} \Pscreen^3}
	{g \, \bar f_{\Pscatter}^2 \Pscatter^3}
    \gsim
	(g^2 \, \bar f_{\Pscatter})^{-1/2}
    \gsim
	O(g^{-\alpha}) \,.
\end {equation}
But the more general condition \ref {item:meff} determines the limit of
applicability of kinetic theory for dilute systems.

In order to regard excitations as having well-defined energies and momenta,
their de Broglie wavelengths must also be small compared to the mean
time between scatterings.
The longest relevant de Broglie wavelength is $1/\Phard$, so the validity
of kinetic theory requires that $\Phard \, \tau_{\rm soft} \gg 1$.
But this automatically follows from
conditions \ref {item:scale-sep} and \ref {item:meff}.


Condition 6, requiring smoothness of distribution functions in
momentum space, prevents applications of our effective theory
to cold degenerate quark matter.
This condition implies that the temperature
(for near-equilibrium systems) must be large compared to
$g \, p_{\rm fermi}$ and hence, for weak coupling, lies far outside
the temperature region in which color superconductivity occurs.

The remainder of this paper is organized as follows.
Section \ref {sec:kinetic theory} presents the structure of
the effective kinetic theory.
The effective scattering amplitudes characterizing $2\lra2$ processes
are discussed in section \ref {sec:2-2}.
These quasiparticle scattering amplitudes 
depend on medium-dependent self-energies,
which are the subject of section \ref {sec:self-energies}.
The appropriate formulation of effective transition rates for
near-collinear ``$1\lra2$" processes is described in section \ref {sec:1-2}.
Section \ref {sec:validity} discusses the validity of our effective
kinetic theory at greater length, including
possible double counting problems in our effective collision terms,
and potential contributions of omitted scattering processes.
We argue that neither of these concerns are an issue.
This section also examines the possible appearance of instabilities
in soft ($p \sim \meff$) modes of the gauge field,
and briefly mentions open problems associated with
extending our effective kinetic theory beyond leading order.
Two short appendices follow.
One summarizes simplifications to the formulas
in the main text that can be made in the case
of isotropic distribution functions,
and the other discusses the connection between the formulas for
$1\lra2$ scattering presented in section \ref {sec:1-2} and the
results for the total gluon emission rate discussed in Ref.~\cite {sansra}.


\section {The effective kinetic theory}\label {sec:kinetic theory}

Our effective kinetic theory
will include all $2\lra2$ processes as
well as effective collinear ``$1\lra2$'' processes.
The Boltzmann equations are,
\begin {equation}
   ( \partial_t + \hat\p\cdot\grad_\x ) \, f_s(\x,\p,t)
   = - C^\twotwo_s[f] - C^{``1\lra2"}_s[f] ,
\label {eq:BOLTZ}
\end {equation}
where the label $s$ denotes the species of excitation
(gluon, up-quark, up-antiquark, down-quark, down-antiquark, {\em etc}.).
Since we have assumed that distributions are not spin or color polarized,
we do not decorate distribution functions with any spin or color label.
However it should be understood that $f_s(\x,\p,t)$ represents the phase
space density of a single helicity and color state of type $s$ quasiparticles.%
\footnote
    {%
    Since our effective theory describes typical hard excitations,
    gluons are to be regarded has having only two transverse polarizations.
    There is a longitudinal branch of the gluon dispersion relation
    in a hot plasma, but the spectral density of longitudinal excitations
    is exponentially small for hard momenta.
    Hence, these collective excitations may be completely ignored
    for our present purposes.
    }

Schematically, the overall structure of our Boltzmann equations is
similar to that outlined by Baier, Mueller, Schiff, and Son \cite{BMSS}
in their treatment of the late stages of their ``bottom-up'' picture of
thermalization in heavy ion collisions, but our formulation
of the details of the collision terms will be guided by our goal
of providing a treatment which is complete at leading order.
As will be discussed below, this requires a consistent treatment
of both screening and LPM suppression of near-collinear processes.

The elastic $2\lra2$ collision term for a given species $a$
has a conventional form,
\begin {align}
   C^\twotwo_a[f]
   ={} & \frac 1{4|\p|\degen_a} \sum_{bcd} \int_{\k\p'\k'}
       \left|{\cal M}^{ab}_{cd}(\p,\k;\p',\k')\strut\right|^2 \,
       (2\pi)^4 \delta^{(4)}(P+K-P'-K')
\nonumber\\ & {} \times
       \Bigl\{
          f_a(\p) \, f_b(\k) \, [1{\pm} f_c(\p')] \, [1{\pm} f_d(\k')]
          - f_c(\p') \, f_d(\k') \, [1{\pm} f_a(\p)] \, [1{\pm} f_b(\k)]
       \Bigl\} \,.
\label{eq:Ctwotwo}
\end {align}
We have introduced $\degen_s$ as the number of spin times color states for
species $s$.
(So
$\degen_s$ equals $6$ for each quark or antiquark, and $16$ for gluons.)
Capital letters denote 4-vectors.
The on-shell 4-momenta appearing inside the delta-function
are to be understood as null vectors,
\begin {equation}
    P^0 \equiv |\p| \,, \qquad \hbox {\em etc.}
\label {eq:massless}
\end {equation}
We are using $\int_\p$ to denote Lorentz invariant momentum integration,
\begin {equation}
    \int_\p \cdots \equiv \int \dprel \cdots \, .
\label {eq:intp}
\end {equation}
The first term in curly braces in (\ref {eq:Ctwotwo})
is a loss term, and the second is a gain term.
${\cal M}^{ab}_{cd}$ denotes an effective scattering amplitude for the process
$ab \lra cd$,
defined with a relativistic normalization for single particle states;
its square, $|{\cal M}^{ab}_{cd}|^2$, should be understood as summed,
not averaged, over the spins and colors of all four excitations
(hence the prefactor of $1/\nu_a$).
The initial factor of $1/(4|\p|)$ is a combination of a final
(or initial) state symmetry factor%
\footnote
    {When the species $c$ and $d$ are identical, a symmetry factor is
    required.  When they are distinct, the final state is double-counted
    in the sum $\sum_{cd}$ over species.
    Hence a factor of $\half$ is needed in either case.}
of $\half$ together with the
$1/(2|\p|)$ from the relativistic normalization of the scattering amplitude.

Symmetry under time-reversal and particle interchange imply that
\begin {equation}
    |{\cal M}^{ab}_{cd}(\p,\k;\p',\k')|^2
    =
    |{\cal M}^{ab}_{dc}(\p,\k;\k',\p')|^2
    =
    |{\cal M}^{ba}_{cd}(\k,\p;\p',\k')|^2
    =
    |{\cal M}^{cd}_{ab}(\p',\k';\p,\k)|^2 \,,
\end {equation}
{\em etc}.
The effective scattering amplitude 
will itself be a functional of the distribution functions,
since the density of other particles in the plasma determines
screening lengths which affect the amplitude for soft scattering.
This will be discussed explicitly in the next section.

As Eq.~(\ref {eq:massless}) makes explicit,
we have neglected medium-dependent corrections to quasiparticle
dispersion relations in the overall kinematics of the collision terms,
and in the particle velocity appearing in the convective derivative on the
left side of the Boltzmann equation (\ref {eq:BOLTZ}).
Given our assumed separation of scales,
these are relative $O(g^{2\alpha})$ perturbations to the energy or velocity
of a hard quasiparticle.
Because we have assumed that distribution functions
and observables are smooth functions on phase space,
including (or excluding) these medium-dependent dispersion relation
corrections will only affect subleading corrections to observables
of interest.
(This is discussed further in sections \ref{sec:2-2} and \ref{sec:validity}.)

Now consider ``$1\lra2$'' processes.
If isolated $1 \lra 2$ processes
are kinematically allowed by the effective thermal
masses of the particles involved,
and if there were no need to consider $1+N \lra 2+N$ processes,
then the appropriate $1\lra2$ collision
term would have a form completely analogous to the $2\lra2$ collision term:
\begin {align}
   C^{1\lra2}_a[f]
   &{}= \frac{1}{4|\p|\degen_a} \sum_{b,c} \int_{\p'\k'}
       \left|{\cal M}^a_{bc}(\p;\p',\k')\strut\right|^2 \,
       (2\pi)^4 \delta^{(4)}(P-P'-K')
\nonumber\\ & \hspace{4em} \times
       \Bigl\{
          f_a(\p) [1\pm f_b(\p')] [1\pm f_c(\k')]
          - f_b(\p') f_c(\k') [1\pm f_a(\p)]
       \Bigr\}
\nonumber\\ & {}
   + \frac{1}{2|\p|\degen_a} \sum_{b,c} \int_{\k\p'}
       \left|{\cal M}_{ab}^{c}(\p';\p,\k)\strut\right|^2 \,
       (2\pi)^4 \delta^{(4)}(P+K-P')
\nonumber\\ & \hspace{4em} \times
       \Bigl\{
          f_a(\p) f_b(\k) [1\pm f_c(\p')]
          - f_c(\p') [1\pm f_a(\p)] [1\pm f_b(\k)]
       \Bigr\} \,.
\label {eq:C12}
\end {align}
With strictly massless kinematics, it is impossible to
satisfy both energy and momentum conservation in a $1\lra2$ particle process
unless all particles are exactly collinear.
For small masses (compared to the energies of the primaries),
the particles will be very nearly collinear.
One could then integrate over the small transverse momenta associated
with the splitting to recast the collision term in the form
\begin {align}
   C^{``1\lra2"}_a[f]
   & {}= 
       \frac{(2\pi)^3}{2|\p|^2 \degen_a}
       \sum_{b,c}
       \int_0^\infty dp' \> dk' \;
       \delta(|\p|-p'-k') \;
       \splitsym^a_{bc}(\p;p'\, \hat \p,k'\, \hat \p)
\nonumber\\ & \hspace{4em} \times
       \Bigl\{
          f_a(\p) [1\pm f_b(p' \, \hat \p)] [1\pm f_c(k' \, \hat\p)]
          - f_b(p' \, \hat\p) f_c(k' \, \hat\p) [1\pm f_a(\p)]
       \Bigr\}
\nonumber\\ & {}
    +
       \frac{(2\pi)^3}{|\p|^2 \degen_a}
       \sum_{b,c}
       \int_0^\infty dk \> dp' \;
       \delta(|\p|+k-p') \;
       \splitsym_{ab}^{c}(p'\, \hat\p;\p,k \,\hat\p)
\nonumber\\ & \hspace{4em} \times
       \Bigl\{
          f_a(\p) f_b(k \, \hat\p) [1\pm f_c(p' \, \hat\p)]
          - f_c(p' \, \hat\p) [1\pm f_a(\p)] [1\pm f_b(k \, \hat\p)]
       \Bigr\} \,,
\label {eq:C12form}
\end {align}
where we have ignored the small deviations from exact collinearity when
evaluating the distribution functions.
The factor of $\splitsym^a_{bc}(\p;p'\hat\p,k\hat\p)$
in the integrand is simply a way of
parameterizing the differential
rate $d\Gamma/dp\,dp'\,dk\,d\Omega_{\hat \p}$ for an $a \to b c$
splitting processes,
integrated over transverse momenta,
excluding distribution
functions and the longitudinal-momentum conserving $\delta$-function.

Any nearly-collinear ``$1\lra2$'' process, now including $1+N \lra 2+N$
soft scattering with emission events,
can be cast into the general form of (\ref{eq:C12form}).  The only
difference is that the differential splitting/joining rates $\splitsym^a_{bc}$
will now implicitly depend on
the distribution functions for the $N$ scatterers.  All of the
phase space integrations for those scatterers, plus the summation over
$N$, will be packaged into $\splitsym^a_{bc}$.
The appropriate values of the splitting rates $\splitsym^a_{bc}$
will be determined simply by requiring that the collision term
(\ref {eq:C12form}) reproduce previous results
in the literature for the rates of ``$1 \lra 2$'' processes.

In the collision term (\ref {eq:C12form}),
we have written the momenta of the splitting
(or joining) particles as though they were exactly collinear and as
though their energy was exactly conserved.
These particles actually receive $O(\meff)$
kicks to their momentum
and energy due to the soft interactions with the other $N$ particles
participating in the near-collinear $1+N \lra 2+N$ process.
As mentioned earlier, this leads to a separation in the directions
of the splitting particles by angles of order $\meff/\Pprimary$.
Treating this process as a strictly collinear $1\lra2$ body process,
and neglecting the soft $O(\meff)$
momentum transfers to other particles in the system,
is an acceptable approximation because of our assumption
that distribution functions (and observables) are smooth functions of momenta.
If this assumption were not satisfied, then we could not factorize the
collision term into a product of distribution functions and
an effective transition rate, as done above.
Given our assumed separation of scales,
the relative error introduced by this idealization is at most $O(g^\alpha)$,
and hence irrelevant in a leading-order treatment.
The differential rates $\gamma^a_{bc}$ are to be understood as summed over
spins and colors of all three participants.
Their explicit form will be discussed in section \ref {sec:1-2}.


\section {\boldmath 2${}\lra{}$2 particle matrix elements}
\label {sec:2-2}

\begin{figure}
\includegraphics[scale=0.30]{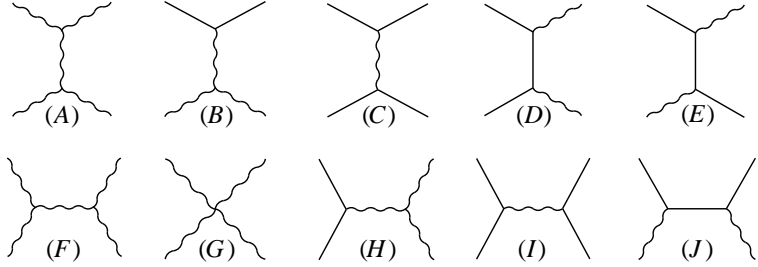}
\caption
    {%
    \label{fig:diagrams}
    Lowest-order diagrams for all $2\leftrightarrow 2$ particle
    scattering processes in a gauge theory with fermions.
    Solid lines denote fermions and wiggly lines are gauge bosons.
    Time may be regarded as running horizontally, either way,
    so a diagram such as $(D)$ represents both
    $q \bar q \to gg$ and $gg \to q \bar q$.%
    }%
\end{figure}

Tree-level diagrams for all $2\lra2$ particle processes in a QCD-like
theory are shown in Figure \ref{fig:diagrams}.
Evaluating these diagrams in vacuum
({\em i.e.}, neglecting medium-dependent self-energy corrections),
squaring the resulting amplitudes, and summing over spins and colors
yields the matrix elements shown in
Table \ref{table:mat_elements}.%
\footnote
    {
    Entries in Table \ref{table:mat_elements} 
    agree with well known QED results and
    the SU(3) results of Combridge {\em et.~al.}~\cite{Combridge}.
    }
Of course, vacuum matrix elements do not correctly describe
the scattering of quasiparticles propagating through a medium.
In principle,
one should recalculate the diagrams of Fig.~\ref{fig:diagrams}
including appropriate medium-dependent self-energy and vertex corrections.
But with generic hard momenta, for which all Mandelstam variables are of
comparable size, these corrections are $O(g^{2\alpha})$ effects
and hence ignorable in a leading-order treatment.%
\footnote
    {%
    Spacetime dependence of distribution functions will not be
    indicated explicitly,
    but it is important to keep in mind our basic assumption that,
    in the vicinity of any point $x$,
    distributions have negligible spacetime variation on
    a scale of $\Tform$.
    All results in this and later sections are local;
    they apply within a region of size $\Tform$ about any
    particular point $x$ of interest, using distribution functions
    evaluated at the point $x$.
    }
Medium-dependent effects can give $O(1)$ corrections to
matrix elements if any Mandelstam variable is $O(\meff^2)$.
Such momentum regions are phase space suppressed by at least
$g^{2\alpha}$ relative to generic hard momenta.
Consequently, momentum regions where $s \sim -t \sim -u \sim \meff^2$
(corresponding to either soft or collinear incoming particles)
give parametrically suppressed contributions,
and medium-dependent corrections can be ignored in these regions.  
This implies that medium-dependent corrections
do not have to be included in terms with denominators of $s^2$.%
\footnote
    {%
    If one were to include self-energy corrections in these $s$-channel
    contributions, then there is a potential subtlety if these terms can
    go on-shell.  This is discussed in section \ref {sec:doublecount},
    but the conclusion that medium-dependent corrections can be ignored
    in $s$-channel processes (when external particles are treated as
    massless) is unchanged.
    }

\begin{table}
\begin{center}
\begin{tabular}{|c|@{\quad}l@{\quad}|}
	\hline &
\\[-5pt]
	$ab \lra cd$ & $\qquad \left|{\cal M}^{ab}_{cd}\right|^2 / g^4$ 
\\[10pt]
	\hline &
\\[-10pt]
	$
	    \begin {array}{c}
		q_1 q_2 \lra q_1 q_2 \,,
		\\ q_1 \bar q_2 \lra q_1 \bar q_2 \,,
		\\ \bar q_1 q_2 \lra \bar q_1 q_2 \,,
		\\ \bar q_1 \bar q_2 \lra \bar q_1 \bar q_2
	    \end {array}
	$
    &
	$
	    \displaystyle
	    8\,  \frac{\df^2 \, \cf^2}{\da}
	    \left( \frac{s^2+u^2}{\ul{t^2}} \right)
	$
\\[30pt]
	$
	\begin {array}{c}
		q_1 q_1\lra q_1 q_1 \,, \\
		\bar q_1 \bar q_1 \lra \bar q_1 \bar q_1 \>
	\end{array}
	$
    & 
	$ \displaystyle
	    8\,  \frac{\df^2 \, \cf^2} {\da} 
	    \left( \frac{s^2+u^2}{\ul{t^2}} + \frac{s^2+t^2}{\ul{u^2}} \right)
	    +
	    16\,  \df \, \cf
	    \left( \cf {-} \frac{\ca}{2} \right) \frac{s^2}{tu}
	$
\\[15pt]
	$q_1 \bar q_1 \lra q_1 \bar q_1$
    &
	$ \displaystyle
	    8\,  \frac{\df^2 \, \cf^2}{\da} 
	    \left( \frac{s^2+u^2}{\ul{t^2}} + \frac{t^2+u^2}{s^2} \right)
	    +
	    16\, \df \, \cf
	    \left( \cf {-} \frac{\ca}{2} \right) \frac{u^2}{st}
	$
\\[15pt]
	$q_1 \bar q_1 \leftrightarrow q_2 \bar q_2$
    &
	$ \displaystyle
	    8\, \frac{\df^2 \, \cf^2}{\da}
	    \left( \frac{t^2 + u^2}{s^2} \right)
	$
\\[15pt]
	$q_1 \bar q_1 \lra g \, g$
    &
	$ \displaystyle
	    8\, \df \, \cf^2
	    \left( \frac{u}{\ul{\ul{t}}} + \frac{t}{\ul{\ul{u}}} \right)
	    -
	    8\, \df \, \cf \, \ca
	    \left( \frac{t^2+u^2}{s^2} \right)
	$
\\[15pt]
	$
	\begin {array}{c}
	    q_1 \, g \lra q_1 \, g \,,\\ \bar q_1 \, g \lra \bar q_1 \, g
	\end {array}
	$
    &
	$ \displaystyle
	    -8\, \df \, \cf^2
	    \left( \frac{u}{s}  +  \frac{s}{\ul{\ul{u}}} \right)
	    +
	    8\, \df \, \cf \, \ca
	    \left( \frac{s^2 + u^2}{\ul{t^2}} \right)
	$
\\[15pt]
	$g \, g \leftrightarrow g \, g$
    &
	$ \displaystyle
	    16\, \da \, \ca^2
	    \left(
		3 - \frac{su}{\ul{t^2}} - \frac{st}{\ul{u^2}} - \frac{tu}{s^2}
	    \right)
	$
\\[10pt]
\hline
\end{tabular}
\end{center}
\caption
    {%
    \label{table:mat_elements}
    Squares of
    vacuum matrix elements for $2\lra2$ particle processes
    in a QCD-like theory, summed over spins and colors
    of all four particles.
    Here $q_1$ and $q_2$ represent quarks of distinct flavors,
    $\bar q_1$ and $\bar q_2$ are the associated antiquarks,
    and $g$ represents a gauge boson.
    $\df$ and $\da$ denote the dimensions of the fundamental
    and adjoint representations, respectively, while $\cf$ and $\ca$ are
    the corresponding quadratic Casimirs.
    In an SU($N$) theory with fundamental representation fermions,
    $\df= \ca = N$, $\cf=(N^2{-}1)/(2N)$, and $\da = N^2{-}1$.
    For a U(1) theory, $\df=\da=\cf=1$ and $\ca=0$.
    For SU(2), $\df = \ca = 2$, $\cf = 3/4$, and $\da = 3$,
    while for SU(3), $\df = \ca = 3$, $\cf = 4/3$, and $\da = 8$.
    Terms with underlined denominators are sufficiently infrared sensitive
    that medium-dependent self-energy corrections must be included,
    as discussed in the text.%
    }
\end{table}

More problematic are the regions where
$-t$ or $-u$ are $O(\meff^2)$ while $s$ is large,
corresponding to a small angle scattering between hard particles.
In this case,
some of the matrix elements in the table are enhanced by
$O(s/\meff^2)$ or $O(s^2/\meff^4)$.
In Table \ref {table:mat_elements}, terms with singly-underlined denominators
indicate such infrared-sensitive contributions arising from
soft gluon exchange,
while terms with double-underlined denominators indicate IR sensitive
contributions from a soft exchanged fermion.
It is these underlined terms in which
medium-dependent effects must be incorporated.%
\footnote
    {%
    Sharp-eyed readers will notice that the $s^2/(tu)$ and $u^2/(st)$ terms
    appearing in the $qq\lra qq$ and $q\bar q \lra q\bar q$
    matrix elements (squared) are not underlined,
    despite the fact that these contributions are infrared-sensitive,
    albeit less so than the $1/t^2$ or $1/u^2$ terms.
    The $s^2/(tu)$ and $u^2/(st)$ terms arise from interference between
    $t$ channel and either $u$ or $s$ channel gluon exchanges.
    They are sufficiently infrared singular to generate
    logarithmically divergent rates in the individual gain
    and loss parts of the collision term,
    but this log divergence (which would be cutoff by medium effects)
    cancels when the gain and loss rates are combined.
    Hence, the apparent IR sensitivity of these terms may be ignored.
    }

For the soft gluon exchange terms one must, in effect,
reevaluate the small $t$ (or small $u$) region of diagrams ($A$), ($B$),
or ($C$) with the free gluon propagator on the internal line replaced by
the appropriate non-equilibrium retarded gluon propagator,
\begin {equation}
    G^{\rm (0)}_{\mu\nu}(Q) =
    \frac {g_{\mu\nu}}{Q^2}
    \longrightarrow
    G^{\rm Ret}_{\mu\nu}(Q)
    \equiv
    \left[ Q^2 + \Pi_{\rm Ret}(Q) \right]^{-1}_{\mu\nu} \,.
\label {eq:Gret}
\end {equation}
(We have chosen Feynman gauge for convenience, but this is not required.)
The required retarded gluon self-energy
$\Pi_{\rm Ret}^{\mu\nu}(Q)$ is discussed in the next section.
Evaluating the propagator (\ref {eq:Gret}) requires, in general,
a non-trivial matrix inversion.
(The matrix inversion may be performed explicitly
in the special case of isotropic distributions;
see Appendix \ref {sec:isotropic}.)

Because the self-energy only matters when the exchange momentum $Q$
is soft, one has considerable freedom in precisely how the
substitution (\ref {eq:Gret}) is implemented.
Different choices, all equally acceptable for a leading-order treatment,
include the following:
\begin {enumerate}
\item
    Fully reevaluate the gluon exchange diagrams ($A$), ($B$), and ($C$),
    using the non-equilibrium propagator (\ref {eq:Gret}) for the internal
    gluon line.
\item
    Introduce a separation scale $\mu$ satisfying $\meff \ll \mu \ll \Phard$,
    and replace the free gluon propagator by the corrected propagator
    (\ref {eq:Gret}) only when $Q^2 < \mu^2$.
\item
    \label {goodchoice}
    Exploit the fact that soft gluon exchange between hard particles
    is spin-independent (to leading order) \cite {sansra}.
    Write the IR sensitive matrix elements as
    the result one would have with fictitious scalar quarks,
    plus an IR insensitive remainder.
    Replace the IR sensitive part by the correct result for scalar
    quarks with medium corrections included.
\end {enumerate}
The final choice is technically the most convenient.
It simply amounts to writing
\begin {eqnarray}
    (s^2 {+} u^2)/{t^2} &=& \half + \half \, (s{-}u)^2/t^2 \,,
\qquad
    s u/t^2 = \quarter - \quarter \, (s{-}u)^2/t^2 \,,
\end {eqnarray}
and then replacing
\begin {eqnarray}
    \frac {(s{-}u)^2}{t^2}
    \longrightarrow
    \left| \strut
	G^{\rm Ret}(P{-}P')_{\mu\nu} \, (P{+}P')^\mu \, (K{+}K')^\nu
    \right|^2 \,.
\end {eqnarray}
To understand this, note that the
square of the vacuum amplitude for $t$-channel gluon exchange between
massless scalars is
\begin {equation}
    \left|
	G^{(0)}(P{-}P')_{\mu\nu} \, (P{+}P')^\mu \, (K{+}K')^\nu
    \right|^2
    =
    \left|\frac {(P{+}P') \cdot (K{+}K')}{(P{-}P')^2} \right|^2
    =
    \frac {(s{-}u)^2}{t^2} \,.
\end {equation}
For the $1/u^2$ terms, apply the same procedure
with $P' \lra K'$, which interchanges $t$ and $u$.

For the soft fermion exchange terms, one must similarly reevaluate
the small $t$ (or $u$) region of diagrams ($D$) and ($E$)
with the internal free fermion propagator replaced
by the non-equilibrium retarded fermion propagator,
\begin {equation}
    \frac {1}{\slashchar Q}
    =
    \frac {\slashchar Q}{Q^2}
    \longrightarrow
    \left[ \slashchar Q - \slashchar \Sigma_{\rm Ret}(Q) \right]^{-1}
    =
    \frac {\slashchar \Q}{\Q^2} \,,
\end {equation}
where $\Q^\mu \equiv Q^\mu - \Sigma_{\rm Ret}^\mu(Q)$.
The retarded fermion self-energy
$\slashchar \Sigma_{\rm Ret}(Q) = \gamma_\mu \Sigma_{\rm Ret}^\mu(Q)$
is discussed in the next section.
In the $q\bar q\lra gg$ matrix element,
the net effect is to replace
\begin {eqnarray}
    \frac {u}{t}
    \longrightarrow
    \frac {4 \, \Re [(P \cdot \Q)(K \cdot \Q)^*] + s \, \Q \cdot \Q^*}
	  {|\Q \cdot \Q|^2}
\end {eqnarray}
with $\Q^\mu = P^\mu{-}P'{}^\mu - \Sigma_{\rm Ret}^\mu(P{-}P')$,
along with the corresponding replacement
with $P' \lra K'$ for the $tu/u^2$ term.
For the $qg \lra qg$ matrix element, the analogous replacement is
\begin {eqnarray}
    \frac {s}{u}
    \longrightarrow
    \frac{-4 \, \Re [(P \cdot \Q)(P' \cdot \Q)^*] 
	+ t \, \Q \cdot \Q^*}
	  {|\Q \cdot \Q|^2}
\end {eqnarray}
with $\Q^\mu$ now equaling $P^\mu {-} {K'}^\mu - \Sigma_{\rm Ret}^\mu(P{-}K')$.
Note that, in general, $|\Q \cdot \Q|^2 \neq (\Q \cdot \Q^*)^2$. 

\section {Soft screening and hard effective masses}\label{sec:self-energies}

To describe correctly the screening effects which cut off
long range Coulomb interactions,
we need the non-equilibrium generalization of the
standard hard thermal loop (HTL) result for the retarded gauge boson
self-energy $\PiR^{\mu\nu}(Q)$ with soft momentum, $Q = O(\meff)$.
The general result for the non-equilibrium case
has been previously derived by \Mrow\ and Thoma \cite{MT},
who obtain%
\footnote{
   The normalization of our distribution functions differ from
   Ref.\ \cite{MT} by a factor of 2,
   which appears in our expression as the spin degeneracy included
   in the factor $\degen_s$.
   To apply (\ref{eq:PiR}) for momenta $Q$ of order $\meff$,
   there is an implicit requirement that the background distributions
   $f(\x,\p,t)$ not vary significantly on time or distance scales of order
   $1/\meff$.
   This is implied by our basic assumption in this paper
   that distributions are smooth on the still longer scale of $\Tform$.
}
\begin {equation}
   \PiR^{\mu\nu}(Q) =
   \sum_s 2 \degen_s \, \frac {g^2 \, C_s}{\da}
       \int \dprel
       \frac{\partial f_s(\p)}{\partial P^\lambda} \left. \left[
       - P^\mu g^{\lambda\nu} + \frac{Q^\lambda P^\mu P^\nu}{P\cdot Q - i\veps}
   \right] \right|_{P^0=|\p|} ,
\label {eq:PiR}
\end {equation}
where $\veps$ is a positive infinitesimal.%
\footnote
    {%
    We use a $({-}{+}{+}{+})$ metric.
    }
In this expression, the
derivative $\partial f(\p)/\partial P^0$ should be understood as zero.
The sum runs over all species of excitations
({\em i.e.}, $g$, $u$, $\bar u$, $d$, $\bar d$, ...),
$\da$ is the dimension of the adjoint representation,
and $C_s$ denotes the quadratic Casimir in the
color representation appropriate for species $s$.%
\footnote
    {
    For Abelian theories, replace $g^2 \, C_s$ by the charge (squared)
    of the corresponding particle.
    }
(See the caption of Table~\ref{table:mat_elements} for specializations.)
The self-energy (\ref{eq:PiR}) can also be rewritten in the form%
\footnote
  {
  To obtain this, replace $d^3\p/2|\p|$ in expression (\ref{eq:PiR}) by
  $d^4P \> \delta(P^2) \, \Theta(P^0)$.  Then integrate by parts, noting
  that the contribution where the derivative hits the delta function
  generates a factor of $P^\lambda$, which vanishes when combined with
  the bracketed expression in (\ref{eq:PiR}).  Then perform the $P^0$ integral
  to get back to $d^3\p$.
  }
\begin {equation}
   \PiR^{\mu\nu}(Q) =
     \sum_s 2 \degen_s \, \frac {g^2 \, C_s}{\da}
     \int \dprel  \, f_s(\p) \left. \left[
         g^{\mu\nu}
         - Q\cdot\partial_P
                    \,\frac{P^\mu P^\nu}{P\cdot Q - i\veps}
      \right] \right|_{P^0=|\p|} \,,
\label {eq:PiR2}
\end {equation}
which is manifestly symmetric in the indices $\mu$ and $\nu$.
Simpler expressions may be given in the special case of isotropic distributions
\cite{CHT},
where $\PiR(Q)$ has the same form
as the equilibrium HTL result, but with a value of
the Debye mass proportional to the integral ${\cal J}$ [Eq.~(\ref {eq:J})]
over the distribution $f(|\p|)$.
This is summarized in Appendix \ref {sec:isotropic}.

We mention in passing that the result (\ref{eq:PiR}) for $\PiR(Q)$
makes the implicit assumption that distributions
$f(\p)$ do not vary significantly for changes of momentum of order $\q$
[which is $O(\meff)$ in the case of interest].  Specifically, the derivation of
(\ref{eq:PiR}) assumes that $f(\p{+}\q) - f(\p)$ can be approximated
as $\q\cdot\grad_\p f(\p)$.

We also need the corresponding self-energy $\SigmaR(Q)$ for fermions with
soft momentum $Q$,
because this cuts off the small momentum-transfer behavior of
fermion exchange diagrams like Fig.\ \ref{fig:qqgg}.
This was also derived by \Mrow\ and Thoma \cite{MT}
but their result has a factor of 4 error.
The corrected expression for a fermion of flavor $s$ is
$\Sigma_{{\rm Ret},s}(Q) = \gamma_\mu \, \Sigma_{{\rm Ret},s}^\mu(Q)$ with
\begin {equation}
   \Sigma_{{\rm Ret},s}^\mu(Q) =
    - g^2 \cf \int \dprel
       \left[\strut 2 f_{\rm g}(\p) + f_s(\p) + f_{\bar s}(\p) \right] \,
       \frac{P^\mu}{P\cdot Q - i\veps} \biggl|_{P^0=|\p|} ,
\label{eq:SigmaR}
\end {equation}
where $f_{\rm g}$, $f_s$, and $f_{\bar s}$ are the gluon, fermion, and
anti-fermion distributions respectively
(as always, per helicity and color state).
The simple gamma-matrix structure $\slashchar \Sigma$
is a consequence of the chiral symmetry which
results from the neglect of zero temperature fermion masses
(relative to their medium-dependent effective mass).

In addition to these soft self-energies, we will also
need medium-dependent corrections to dispersion relations
for hard particles that are nearly on-shell.
These will enter in the formulas that determine the
near-collinear ``$1\lra2$'' rates.
As noted in the introduction,
the resulting dispersion relation corrections for hard excitations,
at leading order, turn out to take the simple form
\begin {equation}
   Q^2 + \meff^2 = 0 \,.
\label {eq:dispersion}
\end {equation}
with a medium-dependent mass $\meff$.
An efficient way to derive the values of these masses is by using the previous
results for soft self-energies, which are valid for $Q \ll \Pscreen$,
in the intermediate regime $\meff \ll Q \ll \Pscreen$,
where both the HTL approximation
and the hard dispersion relation (\ref{eq:dispersion}) are valid.%
\footnote{
   Some readers may wonder whether there can be $O(Q/\Pscreen)$ effects in the
   self-energy that were dropped in the HTL approximation but which
   affect the hard dispersion relation for $Q \gtrsim \Pscreen$,
   and which would cause $\meff$ to be a non-trivial function
   of $Q/\Pscreen$ instead of a constant.
   One can check by explicit
   diagrammatic analysis of self-energies that, for $Q^2 = 0$,
   this does not happen
   for fermions, scalars, or transverse gauge bosons.
%
}

Consider hard gauge bosons first.  Only transverse
polarizations are relevant because the longitudinal
polarization decouples for hard momenta.
We can use the massless dispersion
relation $Q^2=0$ when evaluating the self-energy (\ref{eq:PiR2}) for
the purpose of obtaining the first correction to the dispersion relation.
Working in light-cone coordinates defined by the direction of $\q$,
the second term in brackets in (\ref{eq:PiR2}) is then proportional to
\begin {equation}
   Q^+ \, \frac{\partial}{\partial P^+}
   \left( \frac{P^\mu P^\nu}{P^- Q^+ - i\veps} \right) .
\end {equation}
For transverse choices of $\mu$ and $\nu$, the derivative vanishes.
So the effective mass $\mg$ of hard (transverse) gauge bosons
comes only from the first term of (\ref{eq:PiR2}), giving
\begin {equation}
   \mg^2 =
   \sum_s 2 \degen_s \, \frac{g^2 \, C_s}{\da} \int \dprel \, f_s(\p) \,.
\label {eq:mg}
\end {equation}

Now consider hard fermions.  The effective Dirac equation is
\begin {equation}
   [\slashchar{Q}-\slashchar\Sigma_{\rm Ret}(Q)] \, \psi = 0 \,.
\label {eq:Dirac}
\end {equation}
Multiplying on the left by another factor of
$(\slashchar{Q}-\slashchar\Sigma_{\rm Ret})$ gives the condition
\begin {equation}
   (Q - \SigmaR)^2 = Q^2 - 2 Q\cdot\SigmaR + \SigmaR^2 = 0 \,,
\end {equation}
so that, to leading non-trivial order, the on-shell dispersion relation
for hard fermions is
\begin {equation}
   Q^2 = 2 \, Q \cdot \SigmaR(Q) \,.
\end {equation}
{}From the result (\ref{eq:SigmaR}) for the self-energy, we immediately find
that the effective mass for a hard fermion of flavor $s$ is given by
\begin {equation}
   \meffs^2 = 
    2 g^2 \cf \int \dprel
       \left[\strut 2 f_{\rm g}(\p) + f_s(\p) + f_{\bar s}(\p) \right] .
\label {eq:mf}
\end {equation}

Note that the hard effective masses (\ref {eq:mg}) and (\ref {eq:mf})
are independent of the direction of the momentum $\q$ of the excitation,
as well as its spin,
even in the presence of general anisotropic distributions $f(\p)$.
This is in contrast to soft screening, since the self-energy
$\PiR^{\mu\nu}(Q)$, given by (\ref {eq:PiR}),
generically depends on the direction of $\q$ if
distributions are not isotropic.
The spin independence and isotropy of these (leading order) effective masses
holds provided only that distribution functions are not themselves
polarized.
[This assumption underlies the HTL results (\ref {eq:PiR})
and (\ref {eq:SigmaR}).]
This is special feature of hot gauge theories;
in a generic anisotropic medium, no symmetry argument prevents
splitting of dispersion relations into different branches depending
on the spin of an excitation.
In other words, a generic anisotropic medium is birefringent.

If the self-energies (\ref {eq:PiR}) or (\ref {eq:SigmaR})
had led to spin-dependent dispersion relations for hard excitations,
then even if distributions were not polarized at some initial time,
the subsequent evolution of excitations through the birefringent medium
would generate spin asymmetries.
For such a system, it would be inconsistent to formulate an effective
theory with spin independent but anisotropic distribution functions.
Fortunately, hot gauge theories do not behave this way.

\section {``\boldmath$1 \lra 2$'' particle processes}\label {sec:1-2}

\subsection {Basic formulas}

The appropriate splitting/joining rates
$\splitsym^a_{bc}$ characterizing near-collinear ``$1\lra2$'' processes 
may be extracted from Ref.\ \cite{sansra}.
Let $\n$ be a unit vector in the direction of propagation
of the splitting (or merging) hard particles, so that
$\p = p \, \n$,
$\p' = p' \, \n$, and
$\k = k \, \n$.
Then the required color and spin-summed effective matrix elements,
consistently incorporating the LPM effect at leading order,
may be expressed as
\begin {subequations}
\label{eq:dgamma}
\begin {eqnarray}
	\splitsym^q_{qg}(p \n; p' \n, k \n)
   &=&
	\splitsym^{\bar q}_{\bar qg}(p \n; p' \n, k \n)
   =
	\frac{{\smash{p'}\vphantom p}^2 + p^2}{{\smash{p'}\vphantom p}^2 \, p^2 \, k^3} \>
	{\cal F}_{\rm q}^\n(p,p',k) \,,
\\
	\splitsym^g_{q\bar q}(p \n; p' \n, k \n)
   &=&
	\frac{k^2 + {\smash{p'}\vphantom p}^2}{k^2 \, {\smash{p'}\vphantom p}^2 \, p^3} \>
	{\cal F}_{\rm q}^\n(k,-p',p) \,,
\\
	\splitsym^g_{gg}(p \n; p' \n, k \n)
   &=&
	\frac{{\smash{p'}\vphantom p}^4 + p^4 + k^4}{{\smash{p'}\vphantom p}^3 \, p^3 \, k^3} \>
	{\cal F}_{\rm g}^\n(p,p',k) \,,
\end {eqnarray}
\end {subequations}
where
\begin {equation}
   {\cal F}_s^\n(p',p,k) \equiv
     \frac{d_s \, C_s \, \alpha}{2(2\pi)^3}\,
     \int \frac{d^2 h}{(2\pi)^2} \; 2\h \cdot \Re\, \F_s^\n(\h;p',p,k)
\label {eq:eoo}
\end {equation}
and $\alpha \equiv g^2/(4\pi)$.
The function $\F_s^\n(\h;p',p,k)$,
for fixed given values of $p'$, $p$, $k$ and $\n$,
depends on a two-dimensional vector $\h$ which is perpendicular to $\n$.
$\F_s^\n$ is the solution to the linear integral equation
\begin {align}
    2 \h
    =
    i \, \delta E(\h;p',p,k) \, \F_s^\n(\h;p',p,k)
        + g^2 \int & \frac{d^4Q}{(2\pi)^4} \; 
	2\pi\,\delta(v_\n \cdot Q) \>
	v_\n^\mu \, v_\n^\nu
	    \Dlangle\strut A_\mu(Q) [A_\nu(Q)]^* \Drangle \,
\nonumber\\ {} \times
	\biggl\{
	     (C_s - \half \ca) & \left[{\F}_s^\n(\h;p',p,k)
                       - {\F}_s^\n(\h{-}k\,\q_\perp;p',p,k) \right]
\nonumber\\ {}
	     + \half \ca & \left[{\F}_s^\n(\h;p',p,k)
                       - {\F}_s^\n(\h{+}p'\q_\perp;p',p,k) \right]
\nonumber\\ {}
	     + \half \ca & \left[{\F}_s^\n(\h;p',p,k)
                       - {\F}_s^\n(\h{-}p\,\q_\perp;p',p,k) \right]
	\biggr\} 
	\,,
\label {eq:foo}
\end {align}
which sums up multiple interactions.
The four-vector $v_\n \equiv (1,\n)$ is a null vector in the direction of $\n$,
and the vector $\q_\perp$ is the part of $\q$ perpendicular to $\n$.
Once again, $d_s$ and $C_s$ are the dimension and quadratic Casimir
of the color representation for species $s$.
The energy difference $\delta E$ is defined as
\begin {equation}
    \delta E(\h;p',p,k)
    =
	\frac{\mg^2}{2k} + \frac{\meffs^2}{2p} - \frac{\meffs^2}{2p'} 
	+ \frac{\h^2}{2p \, k\, p'}
\label {eq:deltaE}
\end {equation}
and represents the energy denominator
$\epsilon_g(\k) + \epsilon_s(\p) - \epsilon_s(\p')$
which appears in a $p' \lra pk$ splitting process.
The variable $\h$ is related to the transverse momentum;
see Ref.~\cite {sansra} for details.
We will discuss momentarily the required correlator
$\dlangle A_\mu(Q) [A_\nu(Q)]^* \drangle$
of the soft gauge field.

Ref.\ \cite{sansra}, from which the above formulas for
$\splitsym^a_{bc}$ were
extracted, culminated in the derivation of the near-collinear ``$1\lra2$''
contribution to the total differential gluon production rate
$d\Gamma_{\rm g}/d^3k$ for hard gluons with momentum $k$.
In order to facilitate comparison
with that reference, we show in Appendix \ref{apx:dGamma}
how to express the near-collinear contribution
to $d\Gamma_{\rm g}/d^3k$ in terms of the $\splitsym^a_{bc}$
differential rates.

\begin{figure}
  \includegraphics[scale=0.40]{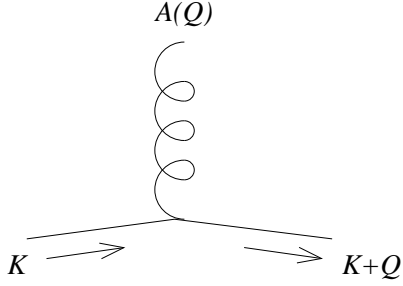}
  \caption{
    Amplitude for the gauge field created by an on-shell hard particle
    in the medium.
    \label{fig:AA}
  }
\end{figure}

The final element we need is the mean square fluctuations in
soft momentum components of the gauge field in the medium,
$\dlangle A_\mu(Q) [A_\nu(Q)]^* \drangle$.
Formally, this is the Fourier transform of the (non-equilibrium)
HTL approximation to the Wightman gauge field correlator,%
\footnote
    {
    Not to be confused with time-ordered or retarded correlators
    of the gauge field.
    Once again,
    our basic assumption is that distribution functions are
    smooth on a time and distance scale of $\Tform$ associated with
    the duration of a near-collinear ``$1\lra2$'' process.
    Hence, for the purpose of evaluating the non-equilibrium correlator
    in (\ref {eq:Wightman}) for momenta of order $\meff$,
    distribution functions may be treated as $x$-independent.
    Spacetime variation in the non-equilibrium state will, of course,
    smear out the momentum conserving delta function in
    (\ref {eq:Wightman}), but only by an amount which is
    irrelevant for our leading order treatment.
    }
excluding the momentum and color conservation delta functions,
\begin {equation}
    \int d^4x \> d^4y \> e^{i (Q'\cdot y - Q \cdot x)} \,
    \Dlangle A_\mu^a(x) \, A_\nu^b(y) \Drangle
    \equiv
    (2\pi)^4 \delta^{(4)}(Q{-}Q') \>
    \delta^{ab}
    \Dlangle \strut A_\mu(Q) [A_\nu(Q)]^* \Drangle \,.
\label {eq:Wightman}
\end {equation}
This correlator characterizes the stochastic background fluctuations
in which collinear splitting processes take place (see Ref.~\cite{sansra}).
We are interested in the correlator for space-like 4-momenta $Q$,
and physically (at leading order) it represents the correlation of the
screened color fields carried by on-shell hard particles streaming randomly
through the plasma.
Hence, it may be understood as the (absolute) square of
the amplitude shown in Fig.\ \ref{fig:AA}
integrated over the phase space of the hard particle, where the gauge
propagator is to be understood as
including the medium-dependent self-energy.
This leads to
\begin {equation}
   \dlangle A_\mu(Q) [A_\nu(Q)]^* \drangle
   = G^{\rm Ret}_{\mu\alpha}(Q) \, \Pi_{12}^{\alpha\beta}(Q)
     \left[ G^{\rm Ret}_{\nu\beta}(Q) \right]^* ,
\end {equation}
where $G^{\rm Ret}(Q)$ is the retarded propagator on the right side
of (\ref{eq:Gret}) and
\begin {equation}
   \Pi_{12}^{\alpha\beta}(Q) \equiv
     \sum_s \degen_s \, \frac {g^2 \, C_s}{\da} \int \frac{d^3\k}{(2\pi)^3} \>
     v_\k^\alpha v_\k^\beta \, f_s(\k) \, [1\pm f_s(\k{+}\q)] \>
     2\pi \, \delta(E_{\k+\q}-E_\k-q^0)
\label {eq:Pi12a}
\end {equation}
for soft momenta $Q$.
Here, $v_\k \equiv (1,\hat \k)$ is the 4-velocity of a hard particle
with momentum $\k$ and $g \, v_\k$ gives the current of this particle 
up to group factors.
The $f$'s appear as initial state distributions
and final-state Bose enhancement or Fermi blocking factors.
For soft momenta $Q$, expression (\ref {eq:Pi12a}) may be
simplified to give
\begin {equation}
   \Pi_{12}^{\alpha\beta}(Q) =
     \sum_s \degen_s \frac {g^2 \, C_s}{\da} \int \frac{d^3\k}{(2\pi)^3} \>
     v_\k^\alpha v_\k^\beta \, f_s(\k) \, [1\pm f_s(\k)] \,
     2\pi\,\delta(v_\k\cdot Q) \,.
\label {eq:Pi12}
\end {equation}
One may alternatively derive this by summing hard thermal loops into
the propagator of the Schwinger-Keldysh formalism, one of whose
components gives the Wightman correlator.%
\footnote{
   For a general introduction, see chapter X of Ref.\ \cite{kinetic}.
   See also, for example, Eqs.\ (A34) and (A23) of Ref.~\cite{ASY}.
}
This is the origin of our notation $\Pi_{12}$ above, which is
the one-loop Wightman current-current correlator and is an off-diagonal
component of the self-energy in this formalism.

In the special case of isotropic distribution functions,
one can simplify substantially the
expression (\ref {eq:Pi12}) for $\Pi_{12}(Q)$.
Moreover, one can
analytically reduce the $d^4 Q$ integral appearing in the
integral equation (\ref {eq:foo}) to a two-dimensional integral
over $\q_\perp$.
See Appendix \ref {sec:isotropic} for details.

In equilibrium, the Wightman correlator is related to the retarded
correlator by the fluctuation-dissipation theorem, which gives
\begin {eqnarray}
   \dlangle A_\mu(Q) [A_\nu(Q)]^* \drangle_{\rm equilibrium}
   &=& 2[n(q^0)+1] \> \Im \, G^{\rm Ret}_{\mu\nu}(Q)
\nonumber\\
   &=& -2[n(q^0)+1] \>
       G^{\rm Ret}_{\mu\alpha}(Q) \, [\Im\,\PiR^{\alpha\beta}(Q)] \,
                      [G^{\rm Ret}_{\nu\beta}(Q)]^* ,
\label {eq:fd}
\end {eqnarray}
where $n(\epsilon) = [e^{\beta\epsilon}{-}1]^{-1}$ is the usual
equilibrium Bose distribution.
For soft $Q$, the prefactor $2[n(q^0)+1]$ can be replaced by $2T/q^0$.
It is instructive to compare the non-equilibrium formula
(\ref{eq:Pi12}) for $\Pi_{12}(Q)$ with that for $\Im\,\PiR(Q)$.
The latter, for soft $Q$, can be extracted from
the earlier expression (\ref{eq:PiR}) which gives
\begin {equation}
   \Im\,\PiR^{\alpha\beta}(Q) =
   \sum_s \degen_s \, \frac{g^2 \, C_s}{\da}
       \int \frac{d^3\k}{(2\pi)^3} \>
       v_\k^\alpha v_\k^\beta \>
       [\q\cdot\grad_\k f_s(\k)] \,
       \pi \, \delta(v_\k\cdot Q).
\label{eq:imPiR}
\end {equation}
The fluctuation-dissipation theorem (\ref{eq:fd}) is satisfied in equilibrium
by the results (\ref{eq:Pi12}) and (\ref{eq:imPiR}) for soft $Q$ because
equilibrium Bose or Fermi distribution functions
$n_s(\epsilon) = [e^{\beta \epsilon}\mp 1]^{-1}$ satisfy
\begin {equation}
   \q\cdot\grad_\k \, n(\epsilon)
   =
   - \beta \, \v\cdot\q \> n_s(\epsilon) \, [1 \pm n_s(\epsilon)] \,.
\end {equation}

\subsection {The formation time}
\label {sec:tform}

Earlier, we promised an explanation of the parametric formulas
(\ref{eq:Tform}) and (\ref{eq:Nform}) for the formation time
$\Tform$ of a near-collinear ``$1\lra2$'' process and for the number
$\Nform$ of soft collisions that take place in that time.
We will give here a brief, superficial review, using the notation
we have adopted in this paper.  For a more thorough discussion of
the scales set by the LPM effect, the reader should consult the
literature, such as Refs.\ \cite{gelis,gyulassy&wang,klein}.
We could deduce the scales by discussing the qualitative
behavior of solutions
of the actual equations (\ref{eq:eoo}) and (\ref{eq:foo})
which incorporate the required physics,
much as we did in the context of photo-emission in Ref.\ \cite{photo_emit}.
Instead, however, we will give here a more physical discussion
which leads to the same results.

One way to understand the basic scales is to begin by
considering classical (soft) \brem\ radiation of photons (rather than
gluons), with wave number $\kgamma$,
emitted by a classical charged
particle moving very close to the speed of light
that undergoes $N$ random small-angle collisions.  (This was the original
classical picture of Landau and Pomeranchuk \cite{LP}.)
Let $\tau$ be the mean time
between those collisions and $\theta_1$ the typical angle of deflection from
each collision.  The total deflection is then $\theta \sim \sqrt N \,\theta_1$,
which we shall assume is small.
If $\tau$ is very large, there will on average be no interference between the
\brem\ fields created by successive collisions (for fixed $\kgamma$).  
If $\tau$ is very small, the \brem\ field will not be able to resolve
the individual collisions, and the field will be the same as that from
a single collision by the total angle $\theta$.  Since \brem\ is at most
logarithmically sensitive to the scattering angle $\theta$, this
smearing of $N$ collisions into one collision will reduce the power
radiated at this wavenumber by
a factor of roughly $N$ compared to what it would be
if each collision could be treated independently.  That is the LPM effect.

The classical \brem\ field from a scattering by angle $\theta$ is dominated
by radiation inside a cone of angle roughly $\theta$.
The first and last scatterings, at space-time points $x_1$ and $x_N$,
will interfere significantly together if the phases in the factors
$\exp(\Kgamma\cdot x_1)$ and $\exp(\Kgamma\cdot x_2)$
are comparable.  For random collisions, that will happen if
$\Kgamma\cdot(x_N{-}x_1) \ll 1$, which, for small $\theta$, gives%
\footnote{
  For sharp single collisions ($N{=}1$), there are actually two angular scales:
  the deflection angle $\theta$ of the charged particle and the
  angle $\theta_\gamma$ that $\k_\gamma$
  makes with the initial or final directions of the particle (whichever
  is smaller).  There can then be logarithms arising from considering
  $\theta_\gamma \ll \theta$ in the \brem\ rate, which we ignore.
}
\begin {equation}
   \kgamma \, N\, \tau \, (1 - \cos \theta)
   \sim \kgamma \, N \, \tau \, \theta^2 \ll 1.
\end {equation}
We have taken
$x_N^0{-}x_1^0 \simeq |\x_N{-}\x_1| \sim N\,\tau$ since the particle
moves on nearly a straight-line trajectory at the speed of light.
For $\kgamma N \tau \theta^2 \gg 1$,
in contrast, the \brem\ produced by the
first and last collisions will be independent.  The crossover criterion
$\kgamma N\tau \theta^2 \sim 1$, with $\theta^2 \sim N \theta_1^2$, then
determines the typical number of collisions which are effectively smeared
together by the LPM effect:
\begin {equation}
   N \sim
   1\Big/\sqrt{\kgamma \tau \theta_1^2} \,,
\label {eq:Nnaive}
\end {equation}
except that $N$ must always be at least 1, since there must be
a scattering to produce classical \brem.

A classical treatment of radiation breaks down when
$\kgamma$ is no longer small compared to the energy $E$
of the radiating charged particle.
However, parametric estimates (as opposed to precise classical formulas)
are still valid where the classical
treatment first begins to break down.  So we can still
use the estimate (\ref{eq:Nnaive}) for $N$ when $\kgamma \sim E$,
provided $E-\kgamma$ is not parametrically small compared to $E$.
In a single soft collision of a hard particle with energy $E$,
the typical deflection angle is
\begin {equation}
   \theta_1 \sim \frac{q_\perp}{E} \,,
\end {equation}
where $q_\perp$ is the typical transverse momentum transfer.
Inserting this into (\ref{eq:Nnaive}) and setting $\kgamma \sim E$ gives
\begin {equation}
   N \sim \max\left(1 , ~
   \sqrt{{E}/(\tau q_\perp^2)}
   \right) .
\label {eq:Nform_est}
\end {equation}
This estimate for hard \brem\ applies equally well to the case of 
gluon emission.
Setting $q_\perp \sim m_\eff$ gives the estimate (\ref{eq:Nform})
previously quoted for $\Nform$.

When the formation time is large compared to $\tau$, then it is simply
$N\tau$ by the definitions of $N$ and $\tau$.
But if $\tau \gg E/q_\perp^2$, so that emission from different
soft collisions do not significantly interfere, then the formation time is
the time scale $t_1$ associated with \brem\ from a single
isolated collision.  Classically, this is the
time $\Delta x^0$ for which $\Kgamma \cdot \Delta x \sim 1$, which gives
$\kgamma t_1 \theta_1^2 \sim 1$ and
\begin {equation}
  t_1 \sim \frac{1}{\kgamma \,\theta_1^2} \sim \frac{E^2}{\kgamma \,q_\perp^2}
  \sim \frac E{q_\perp^2} \,,
\label {eq:t1}
\end {equation}
for $\kgamma \sim E$.
This same result can be found by examining how far off-shell in energy
the internal hard line is in the basic \brem\ processes of
Fig.\ \ref{fig:brem}.
One can nicely combine both cases in the single formula
\begin {equation}
   \Tform \sim \frac{E}{N q_\perp^2} \,,
\label {eq:Tform_est}
\end {equation}
which, upon taking $q_\perp \sim \meff$, gives the earlier quoted result
(\ref{eq:Tform}).%
\footnote
    {
    Replacing $q_\perp$ by $\meff$ in the estimates (\ref {eq:Nform_est})
    and (\ref {eq:Tform_est}) amounts to an implicit assumption that the only
    important collisions are soft collisions with momentum transfers of order
    $\meff$.  Harder collisions are rarer than soft collisions.
    However, as noted in footnote \ref {fn:logs},
    a single collision by an angle $\theta \sim \sqrt{N} \, \meff/E$
    is no rarer than $N$ consecutive soft collisions, each by angle $\meff/E$.
    For large $N$, the same could be said of, for example,
    $N/10$ consecutive collisions, each by angle $\sqrt{10} \, \meff/E$.
    This multiplicity of possibilities turns out to result in logarithmic
    corrections to the above analysis.  Throughout this paper, we have
    consistently ignored logarithms in parametric estimates,
    and we continue to do so here.
    We have also ignored the effect of effective thermal masses, which
    cause hard particles to move slightly slower than the speed of light.
    For $q_\perp \gtrsim \meff$, however, this does not affect any of our
    parametric estimates.
    }

We specialized to $\kgamma \sim E$ above.
In this regime, the above estimates are equally applicable to photon
or gluon emission.
The behavior of the formation time for $\kgamma \ll E$ is not critical
to understanding the conditions for applying our effective theory to
the evolution of hard primaries.  The dominant energy loss mechanism
for primaries is via hard gluon emission processes with $k_{\rm g} \sim E$
rather than $k_{\rm g} \ll E$ soft emissions.  However, it is
interesting to note that for soft emission, the case of photon emission
is qualitatively different from gluon emission.
Eqs.~(\ref {eq:Nnaive}) and (\ref {eq:t1}) show that
the formation time for photons is much longer for $\kgamma \ll E$
than it is for $\kgamma \sim E$.
This is not true for gluon emission.
Because the gluon can scatter by strong interactions, it cannot maintain
its coherence over as long a time scale as a photon can.
Soft scatterings involving the emitted gluon can change its direction
by angles of order $q_\perp/k_{\rm g} \sim \meff/k_{\rm g}$,
which increase as $k_{\rm g}$ decreases.
Consequently, gluon \brem\ with
$k_{\rm g} \ll E$ is actually less coherent than gluon
\brem\ with $k_{\rm g}$ of order $E/2$.  For gluon emission where
the LPM effect is significant, the longest formation time is for the
case where the energies of the two final particles are comparable.


\section {Discussion}\label {sec:validity}

\subsection {Dispersion relation corrections and double counting}
\label {sec:doublecount}

In section \ref {sec:2-2} we asserted that medium-dependent
dispersion relation corrections on external lines are,
for hard particles, sub-leading corrections which may be neglected.
There is, however, a potential subtlety concerning whether or not
the internal line in a $2\lra2$ process is kinematically
allowed to go on-shell.
If this occurs, then the $2\lra 2$ particle collision rate will
include the contribution from an on-shell
$2 \to 1$ process followed by a subsequent $1 \to 2$ process.
Given the presence of explicit $1 \lra 2$ particle collision terms in
our effective theory, this would be inappropriate double-counting
of the underlying scattering events.

\begin{figure}
  \includegraphics[scale=0.70]{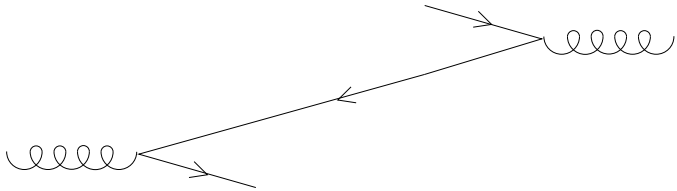}
  \caption{
    The $t$-channel diagram for $q g \to q g$, drawn as a time-ordered
    diagram with time running from left to right.
    \label{fig:factorize}
  }
\end{figure}

Consider, for example, the $t$-channel $qg \to qg$
process illustrated in Fig.\ \ref{fig:factorize}.
To incorporate a correct treatment of small angle scattering,
as discussed in section \ref {sec:2-2},
one must include the HTL fermion self-energy (\ref {eq:SigmaR})
on the internal quark line.
If medium-dependent effective masses are included on the external lines,
then the internal quark line can go on-shell if $\mg > 2\mq$.
This condition is not satisfied in equilibrium QCD,
but it can be satisfied with non-equilibrium distributions.
It can also be satisfied, in equilibrium, in certain QCD-like theories.%
\footnote
    {%
    For example, equilibrium
    SU(3) gauge theories with $\Nf$ Dirac fermions
    have $\mg = (\frac{1}{2} + \frac{1}{12}\Nf)^{1/2} \, g T$ and
    $\mq = \frac{1}{\sqrt3} \, g T$.
    Hence $\mg > 2\mq$ if $\Nf > 10$.
    }
If this condition is satisfied, then the on-shell pole
in the fermion propagator will generate a divergence
in the two body collision term $C^{2\lra2}$.
This divergence arises from an integration over
the time difference between the creation and destruction of the
virtual intermediate fermion, and reflects the fact
that a calculation which just includes the HTL self-energy (\ref {eq:SigmaR})
is modeling that excitation as having an infinite lifetime.%
\footnote
    {%
    This is because the HTL self-energies
    (\ref {eq:PiR}) and (\ref {eq:SigmaR})
    have no imaginary part for timelike 4-momenta~$Q$.
    }
This divergence is, of course, unphysical and would effectively be replaced
by the transport mean free time if further interactions with the medium
were properly included.%
\footnote
    {%
    One natural sounding but inadequate solution is to include the
    full thermal width on the intermediate line.
    However, this width is logarithmically IR divergent
    (in perturbation theory) due to sensitivity to
    long wavelength non-perturbative fluctuations in the gauge field.
    Moreover, merely including a width without simultaneously
    including additional interactions with the medium
    incorporates the wrong physics;
    the width is dominated by soft scattering, but a soft scattering
    event does not prevent the on-shell intermediate particle from
    propagating a large distance before breaking up.
    }

\begin{figure}
  \includegraphics[scale=0.5]{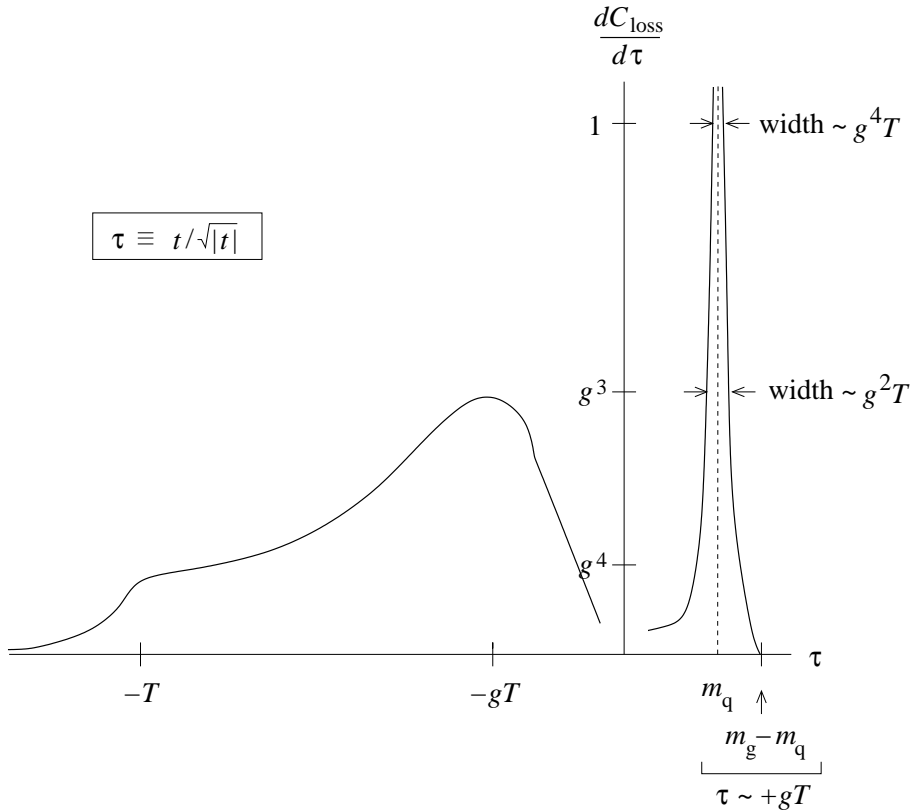}
  \caption{
    A qualitative sketch of the purely $t$-channel
    contribution of Fig.\ \ref{fig:factorize} for $qg \lra qg$ to
    the loss part of the $2\lra2$ collision term
    for a hard gluon or quark,
    plotted as $dC^\twotwo_{\rm loss}/d\tau$ vs.\ $\tau$
    where $\tau\equiv t/\sqrt{|t|}$.
    Interferences with other channels are ignored.
    The labels $m_{\rm q}$ and $m_{\rm g}$ are
    short-hand for the hard effective quark and gluon masses $\mq$ and $\mg$,
    and we have assumed $\mg > 2\mq$.
    The other scales listed above ($1$, $g^4$, $T$, etc.)
    only denote parametric orders.
    \label{fig:CvsT}
  }
\end{figure}

This situation is depicted in more detail (but still qualitatively)
in Fig.\ \ref{fig:CvsT},
which shows the contribution to the loss part of the collision term $C^\twotwo$
(for hard particles) from $t$-channel $qg \lra qg$ scattering,
as a function of the invariant momentum transfer.
For simplicity of presentation, this figure assumes that
the momenta of scatterers, screeners and primaries
are all within an $O(1)$ factor of a common scale $T$,
and that the phase space distributions of all excitations on these scales
are $O(1)$.
In other words, the system is at most an $O(1)$ deviation from equilibrium.
We have found it convenient to use
$
   \tau \equiv t/\sqrt{|t|}
$
rather than $t$, and have sketched $dC^\twotwo_{\rm loss}/d\tau$
vs. $\tau$.
If the external particles were massless,
then $t$ (and hence $\tau$) would always be negative.
The behavior between $\tau \sim -T$ and $\tau \sim -gT$ is
$dC/d\tau \sim g^4/\tau$ and is responsible for a leading-log contribution to
the collision term.
The (exaggerated) fall-off depicted for $\tau \ll -T$
reflects the decrease of initial state distributions for momenta large
compared to $T$.
The fall-off just above $\tau \sim -gT$ is due to screening
(that is, due to the inclusion of the HTL self-energy for the
internal quark line).
For $\tau \lsim -gT$,
the contributions to $C^\twotwo_{\rm loss}$ are
dominated by hard initial particles whose trajectories intersect at an angle
$\theta_{12}$ of order 1.
Each particle is deflected in the collision by an
angle of order $|\tau|/T$.
The contributions in this kinematic region are
not sensitive, at leading order, to whether or not one uses massive or
massless dispersion relations for the external lines.
In contrast, for $\tau > 0$,
the
contributions are dominated by hard initial particles that are nearly
collinear, with $\theta_{12} \sim g$.
The peak at $\tau=\mq$
represents the nearly-collinear process of Fig.~\ref{fig:factorize}.
[Kinematics forces this process to be nearly collinear because the $O(gT)$
masses $\mq$ and $\mg$ are small compared to the hard $O(T)$ momenta.]

   From the plot, one can see that two regions make
significant contributions to the collision term.
The first region is $-T \lesssim \tau \lesssim gT$,
which reflects genuine $2 \lra 2$ processes and makes an
$O(g^4 T)$ contribution to $C_{\rm loss}$.
The second region is $|\tau-\mq| \lesssim g^4 T$,
which is double counting and incorrectly treating the
$1 \lra 2$ processes that are described by the ``$1\lra2$'' collision term.
Fortunately, our treatment of ``$1 \lra 2$'' processes correctly handles
off-shellness in energy as large as the inverse formation time $O(g^2 T)$
(see section \ref{sec:scales})
and so already correctly
accounts for the physics in this region where the energy is off-shell
by only $\lsim g^4 T$.
Note that $\tau$'s further out on the peak at $\mq$ in Fig.\
\ref{fig:tchannel} (for instance, $|\tau-\mq| \sim g^2 T$),
give a contribution to $C_{\rm loss}$ from this particular diagram
that is subleading compared to
genuine $2\lra2$ contributions and which may therefore be ignored.

To formulate a correct collision term which only includes ``genuine''
$2\lra2$ processes, one must somehow keep the contribution from the
first region and eliminate the second.
Although there are many ways one could accomplish this,
the simplest solution is to include
the HTL self-energy on internal lines
(where it is needed to describe screening)
but to treat external lines as massless.

\begin{figure}
  \includegraphics[scale=0.70]{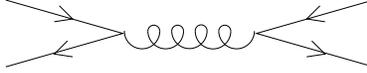}
  \caption{
    $s$-channel diagram for $q \bar q\to q \bar q$.
    \label{fig:schannel}
  }
\end{figure}

\begin{figure}
  \includegraphics[scale=0.5]{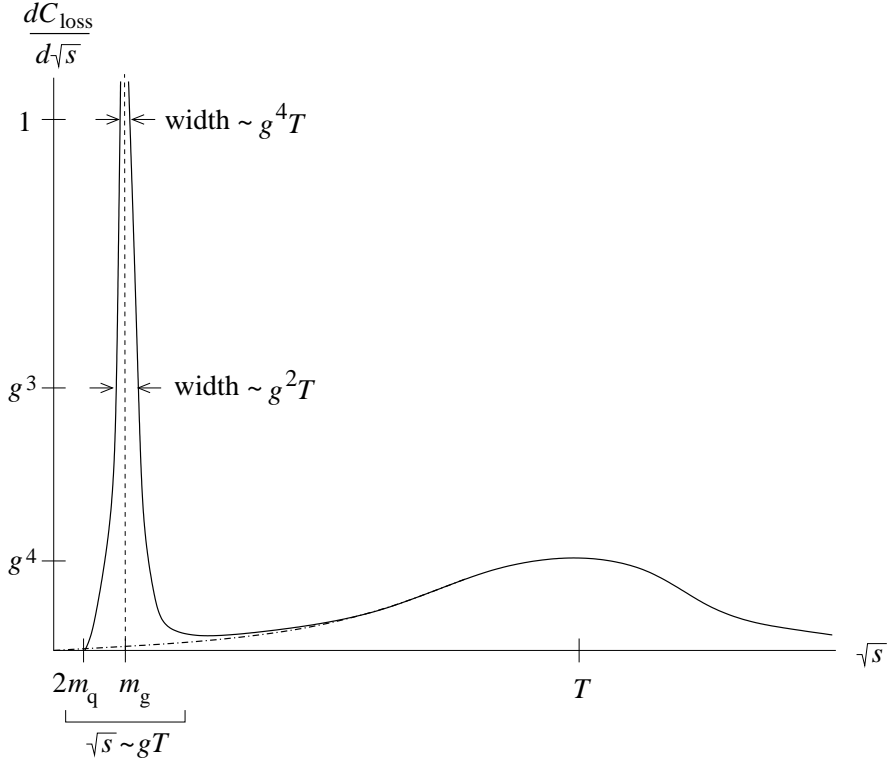}
  \caption{
    A qualitative sketch of the purely $s$-channel
    contribution of Fig.\ \ref{fig:schannel} for $q\bar q \lra q\bar q$
    to
    the collision term $C$ for a hard quark, plotted as $dC/d\sqrt{s}$
    vs.\ $\sqrt{s}$.
    The dot-dash line shows the dependence if $\mq$ and $\mg$
    are both set to zero.
    \label{fig:CvsS}
  }
\end{figure}

Exactly the same issue can arise with $s$-channel processes,
as illustrated in Fig.\ \ref{fig:schannel} for $q\bar q \lra q\bar q$.
A plot analogous to the one discussed above, but this time showing the
contribution to $dC^\twotwo_{\rm loss}/d\sqrt{s}$ vs. $\sqrt{s}$
for this $s$-channel annihilation reaction,
is shown in Fig.\ \ref{fig:CvsS}.
If HTL self-energies are included in the internal gluon line,
then the virtual gluon can go on-shell in this
$q \bar q \to g \to q \bar q$ process
even when the external quarks are treated as massless,
since two massless particles are kinematically allowed to combine to
create one massive one.
For $s$-channel processes, however, it is not necessary to include
HTL self-energies on the internal line in the first place.
Unlike the case of $t$ and $u$-channel processes, they are not
required to control infrared sensitivity.
So one solution which avoids all
partial double counting of $1\lra2$ processes while not affecting the
leading-order result for the true $2\lra2$ contributions is to
treat all external particles as massless, and include HTL
self-energies in $t$-channel and $u$-channel internal propagators but
not in $s$-channel ones.
This is the approach presented in section \ref {sec:2-2}.
The dot-dash line hiding under the peak in Fig. \ref {fig:CvsS}
and smoothly decreasing at small $s$
illustrates the result of this prescription for $s$-channel processes.

There are alternative possibilities which are equally valid at leading order.
For example, one could include HTL self-energies
multiplied by the step function $\Theta(Q^2)$
on all internal lines in $2\lra2$ processes,
so that they only affect spacelike propagators.
With external particles treated as massless, this would also
avoid double-counting mistakes.
But attempting to ``improve'' the effective theory by including
both medium-dependent self-energies on internal lines
and dispersion corrections on external lines is simply wrong ---
unless the contributions from $2\lra2$ processes degenerating into
two independent scatterings are carefully separated and subtracted.
Although this could be done consistently,%
\footnote
    {
    Exactly the same issue of $2\lra2$ scattering processes
    degenerating into two independent $1\lra2$ scatterings
    arises anytime one attempts to
    formulate a kinetic theory for unstable particles.
    Ref.~\cite{KolbWolfram}, for example,
    includes a discussion of the need to subtract
    the degenerating part of $2\lra2$ scattering rates in
    the context of various models of GUT-scale baryogenesis.
    }
it is needlessly
complicated compared to the simple approach of
treating external lines as massless and only inserting
medium-dependent self-energies where they are truly required.

\subsection {Additional scattering processes}
\label {sec:addl-processes}

\begin{figure}
  \includegraphics[scale=0.60]{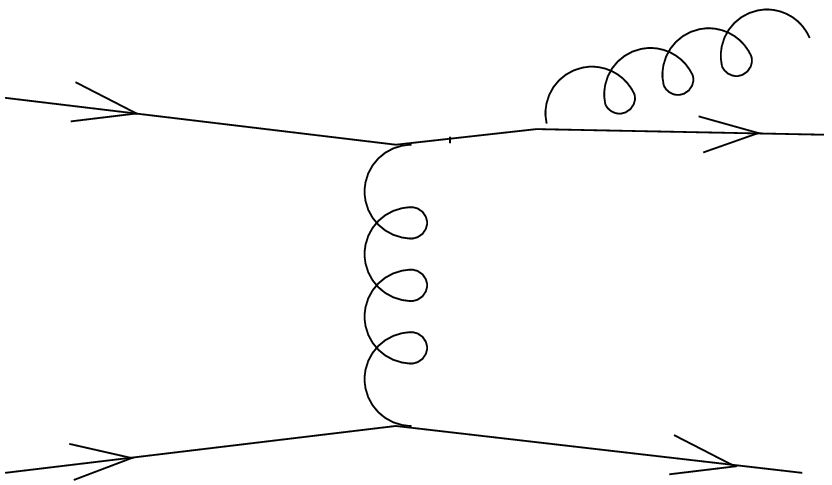}
  \caption{
    A subleading hard $qq \to qqg$ process.
    \label{fig:qq-qqg}
  }
\end{figure}

One may wonder if any additional scattering processes
need to be included in a leading-order effective theory.
To examine this,
first note that processes whose rates are parametrically slower
than hard $2\lra2$ or $1\lra2$ processes will have negligible
({\em i.e.}, subleading in $g$)
effect on the dynamics of a quasiparticle.
For example, consider adding an additional radiated gluon
to a {\it hard}\/ ({\it i.e.}, large momentum transfer)
$2\lra2$ scattering process.
The case of $qq \to qqg$ is illustrated in Fig.~\ref {fig:qq-qqg}.
In the context of high-energy collisions in vacuum,
it is well-known that \brem\ gluons cost a factor of $g^2$ times logarithms
associated with collinear and soft infrared enhancements.
Soft gluon emission is unimportant for our leading-order effective theory
which only describes the dynamics of hard quasiparticles.
And in any case, sensitivity to small gluon momenta
will be cut off in a medium by the effective mass of the gluon.
Collinear logarithms will
similarly be cut off by the effective masses of the quark and gluon.
However, in a medium there is also a $[1{+}f_{\rm g}]$ final state
statistical factor associated with the radiated gluon.
This factor can be parametrically large.
But, as discussed in section \ref {sec:non-eq},
our assumptions limit distribution functions to be
small compared to $O(g^{-2})$ for $p \gsim \meff$.
The upshot is that each radiated gluon suppresses
the transition rate by a factor 
that is parametrically small.
[If we focus on \brem\ of hard gluons, then condition (\ref{eq:strongf})
implies the suppression is at least $g^{2\alpha}$, possibly times
logs of $1/g$.]
Therefore, a primary quark or gluon will experience a parametrically
large number of hard scatterings without gluon \brem\ before it
undergoes one with \brem.

The simple fact that a process is suppressed compared to others
does not automatically make it irrelevant at leading order.  For instance,
$2 \lra 2$ scattering does not change the total number of hard quasiparticles.
Correctly including the dominant number changing processes,
even if they are
parametrically slow compared to the large angle scattering time, is
necessary for a leading-order calculation
of physics that depends on equilibration in the number of quasiparticles
(such as bulk viscosity).%
\footnote
    {
    This is discussed at some length, in the context of bulk viscosity
    for scalar theory, in Refs.~\cite{Jeon,JeonYaffe}.
    }
However, \brem\ from hard $2 \lra 2$ scattering is not the fastest
number changing process.

The most likely scattering events
(as discussed in section \ref {sec:scales})
are soft scatterings with momentum transfer of order $\meff$.
The relative suppression for radiated \brem\ gluons is the same
as above (up to logarithms) and is at least $g^{2\alpha}$ for
hard \brem\ gluons.
That is, an excitation will experience
a parametrically large number of soft scatterings unaccompanied by
hard gluon \brem\ before it undergoes a soft scattering with \brem.
And soft scattering with single \brem\
is the fastest number-changing process (illustrated in Fig.~\ref {fig:brem})
for hard particles.
Fortunately, this process
has already been included in our effective kinetic theory.
It is part of the $N+1 \to N+2$ processes which have been summed
up in our effective ``$1\lra2$'' near-collinear transition rates.%
\footnote
    {
    Recall that
    bremsstrahlung gluons are dominated by angles less than or order of the
    deflection angle in the underlying $2 \to 2$ scattering event.
    Diagrammatically, this occurs because of cancellations between different
    diagrams, such as those of Fig.\ \ref{fig:brem}.
    But it can also be seen by reviewing classical formulas for the
    intensity of \brem\ radiation.
    Since the momentum transfer of the most prevalent collisions
    is order $\meff$, the deflection angle is order $\meff/\Pprimary$
    for a primary excitation.
    In local equilibrium settings, this is order $g$.
    \label {fn:brem-angle}
    }

The basic point is that a hard scattering accompanied by \brem\ does
not change the distributions of quarks or gluons in any way which is
distinct from the much more rapid effects of hard $2\lra2$ scatterings
together with soft scattering accompanied by gluon \brem.
In contrast, a soft scattering accompanied by gluon \brem\ cannot
be neglected, both because it changes particle number and
because a single scattering of this type can produce an $O(1)$
change in the momentum of an excitation
(unlike the more rapid $2\lra2$ soft scatterings)
at a rate which can be competitive with hard $2\lra2$ scattering.

To make the above discussion more concrete, let us briefly specialize
to typical excitations in near-equilibrium systems.
In that case,
the rate of hard scattering with hard gluon \brem\ is $O(g^6 T)$,
which is $O(g^2)$ suppressed relative to the $O(g^4 T)$ rate
of hard $2\lra2$ processes.
The rate of soft scattering with hard \brem\ is $O(g^4 T)$,
which is $O(g^2)$ suppressed relative to the $O(g^2 T)$ rate
of straight $2\lra2$ soft scattering.

\begin{figure}
  \includegraphics[scale=0.60]{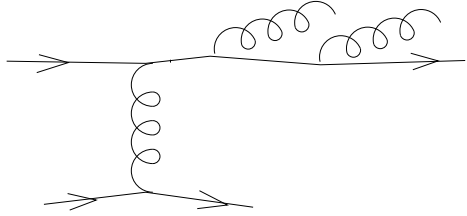}
  \caption{
    A diagram contributing to double gluon \brem.
    \label{fig:brem2}
  }
\end{figure}

As a further example of the suppression of higher-order
processes, consider soft $2 \lra 2$ scattering with hard double
\brem, as depicted in Fig.~\ref{fig:brem2}, which we might call a
``$1\lra3$'' splitting.  As already discussed,
this would be suppressed by $g^{2\alpha}$ compared to single \brem.  It is
therefore irrelevant at leading order,
since a parametrically large number of ``$1\lra2$'' single \brem\
events will occur for every one of these double \brem\ events.
[Near equilibrium, the rate for this ``$1\lra3$'' double \brem\ is $g^6 T$,
compared to the $g^4 T$ rate of ``$1\lra2$'' processes.]

\begin{figure}
  \includegraphics[scale=0.60]{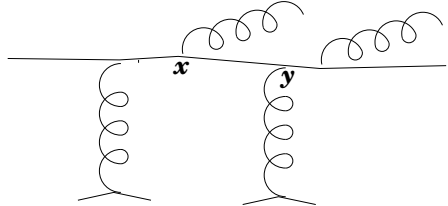}
  \caption{
    Double gluon \brem\ with additional soft scattering.
    \label{fig:3to5}
  }
\end{figure}

Now consider adding one or more additional scatterings to the
double \brem\ process, to generate a diagram like the one shown in
Fig.\ \ref{fig:3to5}.
Adding additional soft scatterings can increase the cross-section, because
the internal line running from $x$ to $y$ in the figure can go on shell.%
\footnote
    {%
    For comparison, the internal quark lines in the
    double \brem\ process of Fig.\ \ref{fig:brem2}
    have virtualities $P^2 \sim \meff^2$ since the angle
    between each hard gluon and the emitting quark line
    is $O(\meff/\Pprimary)$.  (See the previous footnote.)
    Hence these internal lines are off-shell in energy
    by an amount of order $P^2/E \sim \meff^2/\Pprimary$
    (or $g^2 T$ in the near-equilibrium case).
    These internal lines are prevented from being more on-shell by the
    medium-dependent effective masses of particles.
    }
This part of the amplitude, however, really represents two consecutive
``$1\lra2$'' processes and so is already accounted for.
The process of Fig.~\ref{fig:3to5} differs significantly
from two consecutive collisions only
when the time between $x$ and $y$ is comparable to the duration
($1/g^2 T$ near equilibrium)
of the individual ``$1\lra2$'' processes.
In momentum space,
that corresponds to the kinematic region where that propagator is
just as off-shell in energy as the intermediate quark line
in the double \brem, single soft scattering case of
Fig.\ \ref{fig:brem2} considered previously.
In other words,
there is no additional on-shell enhancement
except when the process degenerates into separate scattering events,
which are already included in the effective kinetic theory.

In summary, multiple gluon \brem\ processes are either indistinguishable
from a sequence of ``$1\lra2$'' and $2\lra2$ processes
or else are parametrically slower
and do not accomplish any relevant relaxation not already
provided by a sequence
of faster processes.
After examining these, and other possibilities,
we are unaware of any processes which can affect reasonable
observables at leading order in $g$, beyond those which have already
been included in our effective theory.%
\footnote
    {%
    This assertion does deserve a few caveats.
    Since QCD exactly conserves the net fermion number in each flavor,
    but weak interactions do not,
    weak interaction collision terms cannot be neglected if one is
    interested in the evolution of flavor asymmetries on sufficiently
    long time scales.
    Similarly, in hot electroweak theory explicit baryon
    production/destruction terms representing the effects
    of non-perturbative baryon number changing processes
    can be relevant on time scales large compared to the
    mean free times of processes discussed in this paper.
    See Ref.~\cite {leadinglog} for more discussion of these points.
    }

\subsection {Soft gauge field instabilities}
\label {sec:instability}

In section \ref{sec:2-2}, we noted that in processes involving
soft gluon exchange, one must include the medium-dependent retarded
self-energy in the intermediate gluon propagator in order to
incorporate the screening (and Landau damping) of long range
gauge field interactions.
For situations involving substantial departures from equilibrium,
there is an important question which has not yet been addressed ---
does the medium-dependent self-energy actually cut off long range
interactions?
In other words, are the resulting $2\lra2$ non-equilibrium
scattering rates well-defined?

The relevant part of the phase space integral
for soft gluon exchange processes involves an integral over
the spacelike momentum transfer of the product of advanced
and retarded non-equilibrium gluon propagators,
\begin {equation}
    \int_{Q \ll \Phard} \!\!\!\!\! d^4Q \> \Theta (Q^2) \>
    \left[ Q^2 + \PiR(Q) \right]^{-1}_{\mu\nu}
    \left[ Q^2 + \PiR(Q)^* \right]^{-1}_{\alpha\beta} \,.
\end {equation}
This is then contracted with factors which, by gauge invariance,
are necessarily transverse to $Q$.
If $\chi^a(Q)$ and $\lambda_a(Q)$ denote the eigenvectors
and associated eigenvalues of $\PiR(Q)$, then the potentially
dangerous part of this integral is the piece involving projections
onto the same transverse eigenvector $\chi^a(Q)$ from both propagators,
\begin {equation}
    \int_{Q \ll \Phard} \!\!\!\!\! d^4Q \> \Theta (Q^2) \>
    \frac
	{
	\chi^a_\mu(Q) \,
	\chi^a_\nu(Q)^* \,
	\chi^a_\alpha(Q) \,
	\chi^a_\beta(Q)^*
	}
    { [ Q^2 + \Re \,\lambda_a(Q) ]^2 + [ \Im \, \lambda_a(Q) ]^2} \,.
\label {eq:sing-int}
\end {equation}
The integrand will be singular if the gluon propagator has poles
at real spacelike momenta, which can potentially happen if
an eigenvalue $\lambda_a(Q)$ is real and negative for
some domain of spacelike momenta.
One may show that a spacelike pole in the gluon propagator
creates a non-integrable singularity in the integral (\ref {eq:sing-int})
and generates a logarithmic divergence in the soft collision rate.
The presence of such a spacelike pole in the non-equilibrium retarded
gluon propagator would imply that the corresponding modes of the
soft gauge field have exponentially growing behavior in time.
To our knoweledge, the first discussion of such instabilities
in the context of QCD plasmas was by \Mrow\ in Ref.~\cite{Mrow_instability}.

In equilibrium, no such instability is possible.
More generally, such instabilities do not appear
if distribution functions are isotropic, but otherwise
arbitrarily far from equilibrium.
This is a consequence of the fact, discussed in Appendix \ref {sec:isotropic},
that if distribution functions are rotationally invariant then the
non-equilibrium HTL self-energies turn out to be proportional
to their equilibrium form.

For anisotropic but parity invariant distributions,
it turns out that spacelike poles are generically present.
To see this, first note that the HTL self-energy (\ref{eq:PiR})
does not depend on the magnitude of $Q$, but only on its
(four-dimensional) direction.
And, as may be seen from Eq.~(\ref {eq:imPiR}),
for any parity invariant distribution
the imaginary part of $\PiR(Q)$ vanishes identically when $q^0 = 0$,
implying that the eigenvalues of the zero frequency self-energy are purely real.
Within the $q^0 = 0$ surface,
if there are directions in which $\Re \, \lambda_a(Q)$ is negative
then there will be singularities in the integral (\ref {eq:sing-int}).
For generic anisotropic but parity invariant distributions,
there are such directions.
One may show that the angular average of the trace of the HTL spatial
self-energy, $\Pi_i^i(Q)$, vanishes at zero frequency,
which means that the angular average of the sum of the two transverse
eigenvalues of the spatial zero-frequency self-energy vanishes.
Hence, if there is any direction in which an eigenvalue of
the static spatial HTL self-energy is positive, then there must be some
direction in which an eigenvalue is negative.
The only way instabilities can be avoided is if the zero-frequency spatial
gluon HTL self-energy vanishes identically,
as it does for isotropic distributions.

If one perturbs a parity invariant distribution by adding a
parity non-invariant component, then a continuity argument
shows that instabilities will still generically be present
for sufficiently small deformations away from the parity invariant case.
[In the presence of an arbitrarily small parity non-invariant perturbation,
$\Im \, \lambda_a(Q)$ must vanish on some surface which is a
small deformation of the $q^0 = 0$ plane.
Within this surface, there must still be directions in which
$\Re \, \lambda_a(Q)$ is negative, since there are such directions
in the absence of the deformation.]
Whether these spacelike poles of the gluon propagator persist
for completely general non-parity invariant, anisotropic distributions
is not yet clear to us.

For a given set of distribution functions, if spacelike poles are present
then the characteristic wave vector (and growth rate) of the associated
soft gauge field instabilities will at most be of order of the
non-equilibrium effective mass $\meff$,
since this is the scale which characterizes the size of
medium-dependent self-energy corrections.
For distribution functions which are parametrically close to isotropy,
the wave vector and growth rate of soft instabilities will be
parametrically smaller than $\meff$.
If any gauge field modes with wave vectors of order $\meff$ are unstable,
then the growth of such instabilities would be expected to lead to
spatial (and temporal) inhomogeneities in distribution functions
on length scales of order $1/\meff$.
Consequently, the growth of such instabilities should lead to a violation
of the spacetime smoothness condition which underlies our
effective kinetic theory.
The physics which cuts off the growth in these instabilities,
and removes the divergence in the soft scattering rate,
can only come from including the effects of spacetime inhomogeneities
inside the evaluation of the effective scattering rate.
This means one can no longer use the derivative expansion which
underlies the effective kinetic theory.

As far as our effective kinetic theory is concerned,
the net result is that its domain of validity is smaller
than anticipated.
We initially required that anisotropies in distribution functions
not be parametrically large.
But it appears that $O(1)$ anisotropies in the excitations responsible
for screening will lead to violations of the
assumed spacetime smoothness condition on a time scale that is shorter
than the minimal time scale $\Tform$ for which the effective theory was designed.
However, parametrically small anisotropies are still acceptable.
An $O(g)$ anisotropy in the distribution of ``screeners''
can only generate instabilities whose wave vectors are
parametrically small compared to $\meff$.
In a leading-order treatment, one may simply excise this
``ultrasoft'' momentum region from the phase space integral
(\ref {eq:sing-int}).
The resulting ambiguity in the effective collision rates will be subleading in $g$,
and our effective kinetic theory should remain valid for the intended
class of observables --- those primarily sensitive to the dynamics of hard
excitations.

Clearly, it would be interesting to study the effects of soft gauge field
instabilities in non-equilibrium systems and try to understand their
influence on physical observables which probe the relevant soft
or very-soft dynamics.
This is a topic for future work.

\subsection {Effective kinetic theory beyond leading order?}

Typical effective theories (such as heavy quark theory,
or non-relativistic QED) can be systematically improved
order-by-order in powers of the ratio of scales which
underlies the effective theory.
Having constructed a leading-order effective kinetic theory
for the dynamics of a hot gauge theory,
it is natural to ask whether one can formulate a beyond-leading-order
kinetic theory which will correctly incorporate relative corrections
suppressed by one or more powers of $g$.
This is an interesting open question.

Consider, for simplicity, the case of systems which differ
from equilibrium by at most $O(1)$, so that all relevant hard momenta
are $O(T)$.
Any attempt to construct an effective theory of hot dynamics beyond
leading order must handle numerous different sources of
subleading corrections.  These include:
\begin {enumerate}\advance\itemsep -3pt
\item
    Kinematic mass corrections of order $\meff^2/\Phard^2 \sim g^2$.
    These are everywhere: in the convective derivative of the
    Boltzmann equation, in the overall kinematics of the collision terms,
    inside the effective $2\lra2$ scattering amplitudes, etc.
    Consistently including such corrections should be feasible,
    but will force one to separate and subtract the
    degenerating parts of scattering rates,
    as discussed in section \ref {sec:doublecount}.

\item
    Contributions from higher order tree processes,
    such as \brem\ from hard scattering.
    Numerous such additional processes appear at $O(g^2)$ and
    would need to be included in the collision terms,
    again with appropriate care to eliminate phase space regions
    where an intermediate line goes on-shell and the process
    separates into multiple scattering events.

\item
    Loop corrections to $2\lra2$ effective scattering amplitudes.
    For hard scattering, these will be order $g^2$ effects,
    but for soft scattering, the relevant loop expansion parameter
    is $g$, not $g^2$.  Also, the HTL approximation to the
    self-energies required in soft exchange processes
    receive $O(g)$ corrections because of kinematic approximations
    made in the HTL results.

\item
    Subleading corrections to effective near-collinear transition rates,
    and a proper treatment of the soft emission region.
    We believe these enter at $O(g)$.
    Evaluating such corrections would require a next-to-leading
    order treatment of LPM suppression.
    This is unknown territory.

\item
    Contributions from soft ($\p \sim gT$) on-shell excitations.
    The size of such contributions depends on the sensitivity
    of observables of interest to soft momenta.
    For observables like fermion current densities or the traceless
    part of the stress tensor
    (whose behaviors determines diffusion constants and shear viscosity),
    soft contributions are suppressed by $O(g^4)$ or more.
    Incorporating soft contributions requires formulating
    a kinetic theory which correctly describes both hard
    (ultrarelativistic) and soft (non-relativistic) excitations.

\item
    Contributions from non-perturbative gauge field dynamics
    on the $g^2 T$ (ultrasoft) scale.  The importance of very small
    angle scattering via ultrasoft gauge
    boson exchange is suppressed, relative to soft exchange, by $g^2$; so
    we expect ultrasoft physics effects to enter as $g^2$ corrections
    to our effective kinetic theory.  This means that
    nonperturbative inputs will be necessary to formulate a
    kinetic theory which correctly describes $O(g^2)$ corrections.
    We have no idea how this could be done in practice.

\item
    Corrections due to the uncertainty in energy of excitations.
    The relative size of such corrections is controlled by the
    inverse of an excitation's energy times its mean free time
    between scatterings.
    For soft scatterings of hard excitations, this is $O(g^2)$.
    Correctly incorporating such quantum corrections to kinetic
    theory is an interesting open problem.

\end {enumerate}

Consistently incorporating $O(g)$ corrections
may well be feasible, but extending the effective kinetic theory
to include $O(g^2)$ effects involves major conceptual challenges
as well as technical difficulty.

\section {Conclusion}

We have argued that quasiparticle dynamics in relativistic plasmas
associated with hot, weakly-coupled gauge theories
(such as QCD at asymptotically high temperature $T$)
can be described by an effective kinetic theory,
valid on sufficiently large time and distance scales.
This effective theory is adequate for performing leading-order
evaluations of observables
(such as transport coefficients and energy loss rates)
which are dominantly sensitive to the dynamics
of typical ultrarelativistic excitations.
In other words, our effective theory neglects effects which
generate relative corrections suppressed by powers of the gauge coupling $g$,
but correctly includes all orders in $1/\log g^{-1}$.
To construct such a leading-order effective theory,
it was necessary to include in the collision term of the kinetic theory
both $2\lra2$ particle scattering processes as well as effective
``$1\lra2$'' collinear splitting and merging processes
which represent the net effect of
nearly collinear \brem\ and pair production/annihilation processes
taking place in the presence of fluctuations in the background
gauge field.

Our effective kinetic theory is applicable not only to
near-equilibrium systems
(relevant for the calculation of transport coefficients),
but also to highly non-equilibrium situations,
provided the distribution functions satisfy the conditions
discussed in section \ref {sec:non-eq} [as amended in section \ref {sec:instability}]
which, in particular,
require that there is a clear separation between
the Debye screening scale and the momenta of typical excitations
of interest,
and that the excitations responsible for screening be close to isotropic.
These conditions can be satisfied in asymptotically high temperature QCD,
where the running coupling $g(T)$ is truly small.
They may also be satisfied at intermediate stages of
collisions between arbitrarily large nuclei at asymptotically
high energies \cite {BMSS}.
What, if any, utility this effective theory has for understanding
real heavy ion collisions at accessible energies is not yet clear.
However, we believe that understanding dynamics in
weakly coupled asymptotic regimes is a necessary and useful
prerequisite to understanding dynamics in more strongly coupled regimes.


\begin{acknowledgments}

We are grateful to Stan \Mrow\ for helpful conversations and for
prodding us to think more carefully
about soft gauge field instabilities.
This work was supported, in part, by the U.S. Department
of Energy under Grant Nos.~DE-FG03-96ER40956
and DE-FG02-97ER41027.

\end{acknowledgments}


\appendix

\section {Isotropic distributions}\label {sec:isotropic}

Isotropic (rotationally invariant) distributions represent a physically
interesting class of problems intermediate between completely general
non-equilibrium systems on one hand, and equilibrium or
near-equilibrium systems on the other.  It is widely appreciated that
substantial simplifications occur for equilibrium systems.  For
isotropic systems, almost as high a level of simplification is possible.
The following results are written in the plasma rest frame, which is
uniquely defined for an isotropic system.

We begin with the retarded fermion self-energy
$\SigmaR(K) = \gamma_\mu \, \SigmaR^\mu(K)$,
for soft 4-momentum $K \equiv (k^0, \k)$
such that $k^0 \sim \sqrt{\k^2} \equiv k \ll \Pscreen$.
With isotropic distributions, this soft self-energy has
exactly the same structure as in equilibrium
\cite{Klimov,Weldon1},
\begin{eqnarray}
\Sigma^0(K) & = & \frac{\mf^2}{4k} \left[
	\ln \left| \frac{k^0{+}k}{k^0{-}k} \right|
	-i \pi \, \Theta(k^2 {-} (k^0)^2) \right] , \\
{\bm \Sigma}^i(K) & = & - \k^i \, \frac{\mf^2}{2k^2} \left\{ 1 - 
	\frac{k^0}{2k}   \left[
	\ln \left| \frac{k^0{+}k}{k^0{-}k} \right|
	-i \pi \, \Theta(k^2 {-} (k^0)^2) \right] \right\} .
\end{eqnarray}
The overall coefficient $\mf^2$ is given by the integral (\ref{eq:mf})
which determines the effective mass for hard fermions.
In the literature, this
self-energy is more conventionally written in terms of $M_{\rm q}^2
\equiv \half \mf^2$, because $M_{\rm q}$ is then the frequency
of oscillation for a $\k=0$ fermion.

The case of the retarded gauge field self-energy $\PiR(K)$ is similar.
As in thermal equilibrium \cite{Klimov2,Weldon2}, the self-energy can be
decomposed into longitudinal and transverse pieces,
\begin{eqnarray}
\PiR^{\mu \nu}(K) & = & \Pi_{\rm T}(K) \, P^{\mu \nu}(K) 
	              + \Pi_{\rm L}(K) \, Q^{\mu \nu}(K) \, , \\
\noalign{\hbox{with}}
P^{\mu \nu}(K) & = & \eta^{\mu \nu} + u^\mu u^\nu
	-\frac{\k^\mu \k^\nu}{k^2} \, , \\
Q^{\mu \nu}(K) & = & \frac{(k^2 u^\mu + k^0 \k^\mu)
	(k^2 u^\nu + k^0 \k^\nu)}{k^2 \, [(k^0)^2-k^2]} \, .
\end{eqnarray}
Here $\k^\mu \equiv K^\mu + u^\mu \, u \cdot K = (0,\k)$ denotes
the part of the 4-momentum $K$ orthogonal to the rest frame 4-velocity $u$.
The projectors $P^{\mu\nu}$, $Q^{\mu\nu}$, and $K^\mu K^\nu/K^2$
are mutually orthogonal and sum to the metric,
$
    P^{\mu \nu} + Q^{\mu \nu} + K^\mu K^\nu/K^2 = \eta^{\mu \nu}
$.  
The transverse and longitudinal self-energies are
\begin{eqnarray}
\Pi_{\rm T}(K) & = & \mg^2 \left\{ \frac{(k^0)^2}{k^2}
	+ \frac{k^0(k^2 {-} (k^0)^2)}{2 k^3}
	\left[ \ln \left| \frac{k^0{+}k}{k^0{-}k} \right|
	-i \pi \, \Theta(k^2 {-} (k^0)^2) \right] \right\} , \\
\Pi_{\rm L}(K) & = & 2 \mg^2 \, \frac{k^2{-}(k^0)^2}{k^2}
	\left\{ 1 - \frac{k^0}{2k} \left[  
	\ln \left| \frac{k^0{+}k}{k^0{-}k} \right|
	-i \pi \, \Theta(k^2 {-} (k^0)^2) \right] \right\} ,
\end{eqnarray}
where $\mg^2$ is the asymptotic gluon mass defined in
Eq.~(\protect\ref{eq:mg}).
Equivalent forms in the literature are more commonly
written in terms of the leading order Debye mass
$\mD^2 = 2 \mg^2$, or occasionally in terms of the leading order plasma
frequency $\omega_{\rm pl}^2 = \mD^2/3$.
Note that the definition of $\Pi_{\rm L}$ is not uniform in the literature
(even in previous work by the authors of the present paper!).
The above notation agrees with that of Weldon \cite{Weldon2}.
The other common usage is that of Braaten and Pisarski \cite{BraatenPisarski},
who define $\Pi_{\rm L}$ to be $- k^2 / K^2$ times our value,
so that
$\Pi_{\rm L}^{\hbox{\tiny (Braaten-Pisarski)}} = -\PiR^{00}$.

Rotational symmetry,
even in the absence of equilibrium,
is sufficient to derive
a relation between the imaginary part of the retarded
self-energy and the Wightman self-energy, at soft momenta.
Namely,
\begin {equation}
    \Pi_{12}^{\alpha\beta}(K)
    =
    - \frac {2 T_*}{k^0} \;
    \Im \, \PiR^{\alpha\beta}(K) \,,
\label {eq:fluc-diss}
\end {equation}
where
\begin{equation}
T_* \equiv \, 
   {\displaystyle \sum_s \degen_s \, \frac{g^2 \, C_s}{\da} 
	\> {\cal I}_s
    }
    \Bigm/
   {\displaystyle \sum_s \degen_s \, \frac{g^2 \, C_s}{\da} 
	\> {\cal J}_s
    }
	 \,,
\end{equation}
and ${\cal I}_s$ and ${\cal J}_s$ are the integrals defined in
Eqs.~(\ref {eq:I}) and (\ref {eq:J}), evaluated with the distribution
function for species $s$.
Relation (\ref {eq:fluc-diss}) is just the small frequency form
of the equilibrium fluctuation-dissipation relation (\ref{eq:fd}),
but with the equilibrium temperature replaced by $T_*$.

Consequently, the Wightman correlation function, for soft momenta,
is merely a rescaled version of the thermal Wightman correlation function,
\begin {equation}
    \dlangle A_\mu(K) [A_\nu(K)]^* \drangle
    \Bigr|_{\genfrac{}{}{0pt}1 {\rm non-eq.}{\rm isotropic}}
    =
    \frac {T_*}{T_{\rm eff}} \>
    \dlangle A_\mu(K) [A_\nu(K)]^* \drangle
    \Bigr|_{\genfrac{}{}{0pt}1 {\rm equil.}{T=T_{\rm eff}}} \,,
\label {eq:isoWight}
\end {equation}
at the temperature $T_{\rm eff}$ for which the equilibrium Debye mass
coincides with the correct effective Debye mass,%
\footnote
    {%
    Explicitly, for an SU($\Nc$) theory with $\Nf$ fundamental Dirac
    fermions,
    $
	\frac 16 (\Nc + \Nf \, \cf \, \df /\da) \, g^2 T_{\rm eff}^2
	=
	\mg^2
    $.
    }
$
    \mD^2(T_{\rm eff})|_{\rm equil.}
    =
    2 \mg^2
$.
Note that the value of $T_*$ is not in general the same as $T_{\rm eff}$.

Relation (\ref {eq:isoWight})
permits one to use recent results of Aurenche
{\it et~al.} \cite {AGZ} to reduce the four dimensional integral
involving the Wightman correlator appearing in the
integral equation (\ref {eq:foo})
down to a two dimensional integral over transverse momenta.
One may show that
\begin{align}
g^2 & \int \frac{d^4Q}{(2\pi)^4} \; 2\pi\,\delta(v_\n \cdot Q) \>
	v_\n^\mu \, v_\n^\nu
	\Dlangle\strut A_\mu(Q) [A_\nu(Q)]^* \Drangle
	\Bigr|_{\genfrac{}{}{0pt}1 {\rm non-eq.}{\rm isotropic}}
    \; h(\q_\perp)
\nonumber\\
	& {} =
	g^2 \, T_* \int \frac{d^2 \q_\perp}{(2\pi)^2}
	\left( \frac{1}{\q_\perp^2} - \frac{1}{\q_\perp^2 {+} \mD^2}
	\right)
    h(\q_\perp)
	\, ,
\end{align}
where $h(\q_\perp)$ is any function of $\q_\perp$.


\section{Relationship of \boldmath$d\Gamma_{\rm g}/d^3k$ to
   $\splitsym^{a}_{bc}$}
\label {apx:dGamma}

The leading-order equilibrium differential rate for production
for hard gluons, as defined in Ref.~\cite{sansra},
corresponds to the gain part of gluon collision terms (\ref {eq:Ctwotwo})
and (\ref {eq:C12form}), evaluated in equilibrium and multiplied by
$\nu_g/(2\pi)^3$.
Explicitly, the near-collinear LPM-suppressed part of the
production rate for hard gluons with momentum $\k = k \, \n$ is,
at leading order,
\begin {align}
   \frac{d\Gamma_{\rm g}^\supLPM}{d^3k}
   =
   \frac{1}{4\pi k^2} \,\frac{d\Gamma_{\rm g}^\supLPM}{dk}
   \quad\> &
\nonumber\\ {}
   =
   \frac{[1{+}n_{\rm b}(k)]}{k^2}
   \Biggl(
     \half & \int_0^\infty \! dp\> dp'\; 
       \delta(p'{+}p{-}k) \;
       \splitsym^g_{gg}(k \n; p' \n, p\n)
       \, n_{\rm b}(p') \, n_{\rm b}(p)
\nonumber\\ {}
     + & \int_0^\infty \! dp\> dp'\;
       \delta(p'{-}p{-}k) \;
       \splitsym^g_{gg}(p' \n; p \n, k\n)
       \, n_{\rm b}(p') \, [1{+}n_{\rm b}(p)]
\nonumber\\[5pt] {}
     + \Nf & \int_0^\infty \! dp\> dp'\;
       \delta(p'{+}p{-}k) \;
       \splitsym^g_{q\bar q}(k \n; p' \n, p\n)
       \, n_{\rm f}(p') \, n_{\rm f}(p)
\nonumber\\ {}
     + 2 \Nf & \int_0^\infty \! dp\> dp'\;
       \delta(p'{-}p{-}k) \;
       \splitsym^q_{qg}(p' \n; p \n, k\n)
       \, n_{\rm f}(p') \, [1{-}n_{\rm f}(p)]
     \Biggr).
\label {eq:B1}
\end {align}
The $\half$ in the first term is an initial-state symmetry factor,
the 2 multiplying the final term reflects the identical contributions
from $q \to qg$ and $\bar q \to \bar q g$,
and $\Nf$ is the number of Dirac fermion flavors.
$n_{\rm b}(\omega)$ and $n_{\rm f}(\omega)$ are equilibrium Bose and Fermi
distribution functions, respectively.
Ref.\ \cite{sansra} expresses results more compactly by using
crossing symmetries.
For a more direct comparison with that paper, expression (\ref {eq:B1})
can be rewritten as
\begin {align}
   \frac{d\Gamma_{\rm g}^\supLPM}{d^3k}
   =
   \frac{[1{+}n_{\rm b}(k)]}{k^2}
   \Biggl(
     \half & \int_{-\infty}^\infty dp\> dp'\;
       \delta(p'{-}p{-}k) \;
       \splitsym^g_{gg}(p' \n; p \n, k\n)
       \, n_{\rm b}(p') \, [1{+}n_{\rm b}(p)]
\nonumber\\ {}
     + \Nf & \int_{-\infty}^\infty dp\> dp'\;
       \delta(p'{-}p{-}k) \;
       \splitsym^q_{qg}(p' \n; p \n, k\n)
       \, n_{\rm f}(p') \, [1{-}n_{\rm f}(p)]
     \Biggr).
\label {eq:final}
\end {align}

However, the authors of Ref.~\cite{sansra} should be profoundly chastised
for not pointing out that this differential gluon production rate is,
in fact, an infrared divergent quantity.
The problem arises from the $p \to 0$ region of the $g \lra gg$
term in (\ref {eq:final}).
This portion of the integral represents processes in which
a hard gluon with momentum $\p'$ nearly equal to $\k$
experiences a soft scattering with emission or absorption
of a soft gluon to yield a hard gluon with momentum $\k$.
But physical quantities can only depend on this production
rate minus the corresponding rate at which gluons are scattered
out of the mode $\k$, and the infrared sensitivity cancels
in the difference of these rates.
In other words, although the production rate ${d\Gamma_{\rm g}^\supLPM}/{d^3k}$
is not actually well-defined,
the complete collision terms (\ref {eq:C12form}) built from the same
near-collinear transition amplitudes are infrared safe.

\newpage


\end {document}